\documentclass[twocolumn,superscriptaddress,showpacs,amsmath,amssymb]{revtex4-2}

\usepackage{amsmath,amssymb,amsfonts}
\usepackage{booktabs}          % professional table lines
\usepackage{graphicx}          % figure inclusion
\usepackage{xcolor}            % colored text / boxes
\usepackage{hyperref}
\hypersetup{
  colorlinks=true,
  linkcolor=blue,
  citecolor=blue,
  urlcolor=blue
}
\usepackage{microtype}         % better text justification
\usepackage{tikz}
\usetikzlibrary{shapes.arrows,positioning,fit,calc}
\usepackage{multirow}
\usepackage{orcidlink}



\newcommand{\qsvm}{\textsc{QSVM}}
\newcommand{\svm}{\textsc{SVM}}
\newcommand{\ket}[1]{|#1\rangle}
\newcommand{\bra}[1]{\langle#1|}

\newcommand{\todo}[1]{}
\newcommand{\pending}[1]{}

\newcommand{\dax}[1]{}
\newcommand{\gst}[1]{}

\newcommand{\qinc}{
Quantum Innovation Centre (Q.InC), Agency for Science, Technology and Research (A*STAR), 2 Fusionopolis Way, Innovis \#08-03, Singapore 138634, Republic of Singapore\looseness=-1}

\newcommand{\sutd}{Science, Mathematics and Technology Cluster, Singapore University of Technology and Design, 8 Somapah Road, Singapore 487372, Republic of Singapore\looseness=-1}

\newcommand{\smu}
{Singapore Management University, 81 Victoria St, Singapore 188065\looseness=-1}

\newcommand{\ihpc}{Institute of High Performance Computing (IHPC), Agency for Science, Technology and Research (A*STAR), 1 Fusionopolis Way, \#16-16 Connexis, Singapore 138632, Republic of Singapore\looseness=-1}

\begin{document}
% ============================================================

\title{Quantum Kernel Advantage over Classical Collapse in Medical Foundation Model Embeddings}

\author{Sebastian Cajas Ord\'o\~nez\,\orcidlink{0000-0003-0579-6178}}
\email{sebasmos@mit.edu}
\affiliation{MIT Critical Data, Massachusetts Institute of Technology, Cambridge, MA, USA}

\author{Felipe Ocampo Osorio\,\orcidlink{0000-0002-5250-4636}}
\affiliation{MIT Critical Data, Massachusetts Institute of Technology, Cambridge, MA, USA}
\affiliation{Clinical Research Center, Artificial Intelligence Unit, Fundaci\'{o}n Valle del Lili, Cali, Valle del Cauca, Colombia}

\author{Dax Enshan Koh\,\orcidlink{0000-0002-8968-591X}}
\affiliation{\qinc}
\affiliation{\ihpc}
\affiliation{\sutd}

\author{Rafi Al Attrach\,\orcidlink{0009-0005-0479-7437}}
\affiliation{MIT Critical Data, Massachusetts Institute of Technology, Cambridge, MA, USA}

\author{Aldo Marzullo\,\orcidlink{0000-0002-9651-7156}}
\affiliation{Department of Electronics, Information and Bioengineering, Politecnico di Milano, Milan, Italy}

\author{Ariel Guerra-Adames\,\orcidlink{0000-0002-7881-8246}}
\affiliation{Bordeaux Population Health Research Center, Inserm U1219, Université de Bordeaux, F-33000, Bordeaux, France}
\affiliation{Inria Bordeaux, Université de Bordeaux, F-33000 Bordeaux, France}

\author{J. Alejandro Andrade\,\orcidlink{0000-0003-1406-3064}}
\affiliation{Universidad del Cauca, Popay\'{a}n, Colombia}

\author{Siong Thye Goh\,\orcidlink{0000-0001-7563-0961}}
\affiliation{\ihpc}
\affiliation{\smu}

\author{Chi-Yu Chen\,\orcidlink{0009-0004-7481-0185}}
\affiliation{National Taiwan University Hospital}

\author{Rahul Gorijavolu\,\orcidlink{0000-0002-4386-957X}}
\affiliation{MIT Critical Data, Massachusetts Institute of Technology, Cambridge, MA, USA}
\affiliation{School of Medicine, Johns Hopkins University, Baltimore, MD, USA}
\affiliation{Department of Biomedical Engineering, Johns Hopkins University, Baltimore, MD, USA}
\affiliation{AI for Responsible, Generalizable, and Open Surgical (ARGOS) Research Group, Baltimore, MD, USA}

\author{Xue Yang\,\orcidlink{0009-0006-8132-2686}}
\affiliation{School of Information Engineering, Shanghai Maritime University, Shanghai, 201306, China}
\affiliation{\qinc}
\affiliation{Research Center of Intelligent Information Processing and Quantum Intelligent Computing, Shanghai, 201306, China}

\author{Noah Dane Hebdon\,\orcidlink{0000-0002-2855-0798}}
\affiliation{\qinc}

\author{Leo Anthony Celi\,\orcidlink{0000-0001-6712-6626}}
\affiliation{MIT Critical Data, Massachusetts Institute of Technology, Cambridge, MA, USA}
\affiliation{Laboratory for Computational Physiology, MIT, Cambridge, MA, USA}
\affiliation{Department of Medicine, Beth Israel Deaconess Medical Center, Boston, MA, USA}

% ============================================================
\begin{abstract}
We provide evidence of quantum kernel advantage under noiseless simulation
in binary insurance classification on MIMIC-CXR chest radiographs using
quantum support vector machines (QSVM) with frozen embeddings from three
medical foundation models (MedSigLIP-448, RAD-DINO, ViT-patch32).
We propose a two-tier fair comparison framework in which both classifiers
receive identical PCA-$q$ features; at Tier~1 (untuned QSVM vs.\ untuned
linear SVM, C\,=\,1 both sides), QSVM wins minority-class F1 in all 18
tested configurations (10 embedding seeds; 17 at $p<0.001$, 1 at
$p<0.01$, paired bootstrap).
The classical linear kernel collapses to majority-class prediction
(F1\,=\,0) on 90--100\% of seeds at every qubit count, while QSVM
maintains non-trivial recall.
At $q=11$ (MedSigLIP-448 plateau center), QSVM achieves mean
F1\,=\,0.343\,$\pm$\,0.170 vs.\ classical F1\,=\,0.050\,$\pm$\,0.159
($\Delta$F1\,=\,$+$0.293, $p<0.001$) without hyperparameter tuning.
Under Tier~2 (untuned QSVM vs.\ C-tuned RBF SVM), QSVM wins all seven
tested configurations (mean gain $+$0.068, max $+$0.112).
Eigenspectrum analysis reveals the mechanism: multi-seed mean quantum
kernel effective rank reaches 69.80 at $q=11$, far exceeding the linear
kernel rank of exactly $q=11$, while classical collapse remains
C-invariant.
At $q=16$, any concentration collapse is seed-dependent: multi-seed mean
F1 is 0.377, a Tier-1 win.
A full qubit sweep reveals architecture-dependent concentration onset
across models.
Code: \url{https://github.com/sebasmos/qml-medimage}.
\end{abstract}

% ============================================================
\maketitle
% ============================================================

% ============================================================
\section{Introduction}
\label{sec:intro}
% ============================================================

Quantum machine learning (QML) promises computational advantages through the
use of quantum feature maps that embed classical data into exponentially large
Hilbert spaces~\cite{havlicek2019,schuld2019}.
Quantum kernel methods, and in particular the quantum support vector
machine (\qsvm{}), realize this promise by computing inner products of
quantum states instead of explicit feature vectors, potentially enabling
richer decision boundaries with fewer parameters than classical
alternatives~\cite{schuld2021,liu2021}.
Despite considerable theoretical interest, empirical demonstrations of quantum
advantage on real-world medical imaging tasks remain rare,
partly because rigorous fair comparisons require careful control of
hyperparameters, dimensionality, and regularization on both the
classical and quantum sides~\cite{jerbi2023,bowles2024}.

This work builds on preliminary results presented at the AIQxQIA~2025
workshop~\cite{cajas2025ecai}, substantially expanding the experimental
scope with a two-tier fair comparison framework, multi-model evaluation,
and mechanistic analysis of the classical kernel collapse phenomenon.
We study binary insurance classification on the MIMIC-CXR chest radiograph
dataset~\cite{johnson2019mimic,johnson2023mimicjpg}, predicting whether a
patient holds Private insurance versus Medicaid/Medicare coverage.
Recent work has shown that deep learning models trained on chest radiographs can predict attributes not visually apparent to clinicians,
including self-reported race~\cite{gichoya2022} and insurance type~\cite{chen2025insurance}, even when images are clinically normal. This phenomenon has been hypothesized to arise from spurious correlations: image features statistically associated with demographic or socioeconomic variables, specific positioning conventions, or markers of cumulative environmental exposure, without any direct causal relationship to pathology~\cite{jones2024causal}. When models capture these latent signals rather than genuine clinical features, performance becomes brittle outside the training distribution and errors concentrate in underrepresented groups \cite{seyyedkalantari2021}. Our objective is to evaluate whether quantum kernels improve separability within this representation space, without claiming the learned signal is clinically causal.
These findings carry direct implications for health equity: if socioeconomic and demographic signals are encoded in medical images, clinical AI systems trained on those images risk learning and perpetuating disparities, a concern supported by evidence that chest X-ray classifiers systematically underdiagnose underserved populations~\cite{seyyedkalantari2021,obermeyer2019}. 

Insurance status is already recorded in the electronic health record, so the goal here is not clinical deployment. Insurance prediction provides a clinically grounded test of whether quantum feature maps can extract discriminative structure that classical kernels miss, on a task whose difficulty (subtle, distributed signal in a class-imbalanced setting) makes the comparison meaningful.

Instead of hand-crafted image features, we extract high-dimensional
embeddings from three frozen medical foundation models
(MedSigLIP-448~\cite{zhai2023siglip}, RAD-DINO~\cite{perezgarcia2024raddino},
ViT-patch32~\cite{dosovitskiy2020vit}),
compress them to $q$ dimensions via PCA, and compare \qsvm{} against
classical \svm{} baselines at identical feature dimensionality.
This setting is representative of realistic small-sample quantum pipelines:
the quantum hardware constraint limits practical training to ${\sim}2{,}000$
samples, which is naturally met by PCA reduction to $q \leq 16$ dimensions.

This paper makes four contributions.

\begin{enumerate}
  \item \textbf{Quantum kernel advantage across all tested configurations.}
    \qsvm{} (C=1, reps=1 [\S\ref{sec:circuit}], trace normalization) beats an equally untuned
    linear \svm{} on minority-class F1 at all 18 model$\times$qubit configurations
    ($q \in \{4,6,8,9,10,11,12,16\}$, three models), validated across 10 embedding seeds
    (17 at $p<0.001$, 1 at $p<0.01$; paired bootstrap).
    Classical linear \svm{} collapses to F1\,=\,0 on 90--100\% of seeds at every qubit count.
    Against the best C-tuned RBF kernel at equal PCA dimensionality, \qsvm{} still wins all
    7 configurations (mean gain $+$0.068).

  \item \textbf{Structural explanation for the classical collapse.}
    PCA-$q$ compression leaves the linear kernel with effective rank equal to $q$
    (3.77--5.85 out of $N\!=\!1{,}896$ training samples), making collapse independent of
    regularization parameter~C.
    The quantum kernel reaches 6.86 and 13.94 at $q=4$ and $q=6$ (seed~0;
    $1.82\times$ and $2.52\times$ the linear values), with the ratio growing with qubit count.
    A 10-seed rank-matched RBF experiment confirms the advantage extends beyond effective rank:
    \qsvm{} outperforms an RBF kernel tuned to the same rank at all four qubit counts tested.

  \item \textbf{Three design rules for quantum kernel pipelines.}
    Trace normalization is necessary for non-zero \qsvm{} F1; Frobenius normalization
    collapses it to zero on all models.
    1-DOF angle encoding (one Ry per qubit) consistently outperforms the 3-DOF variant
    (Rz-Ry-Rz).
    Increasing re-uploading depth at $q=8$ degrades performance; the bottleneck is
    sample size rather than circuit capacity.

  \item \textbf{Architecture-dependent concentration.}
    A sweep over $q \in \{2,\ldots,16\}$ reveals model-specific behaviour: on seed~0,
    MedSigLIP-448 peaks at $q=11$ then collapses at $q=16$ (multi-seed mean 0.377,
    a Tier-1 win), while RAD-DINO and ViT-patch32 improve monotonically.
    The variation is consistent with data-dependent concentration rates described by
    Thanasilp et al.~\cite{thanasilp2022} and extends those findings to frozen medical
    foundation model embeddings.
\end{enumerate}

The paper is organized as follows.
Related work (\S\,\ref{sec:related}) covers quantum kernel methods, medical
foundation models, and quantum advantage benchmarking.
The methods (\S\,\ref{sec:methods}) cover the dataset, preprocessing,
circuit design, kernel computation, and fair comparison framework.
Results (\S\,\ref{sec:results}) report the main experiments; ablations
(\S\,\ref{sec:ablation}) address normalization, qubit count, circuit
depth, and data-type variants.
The discussion (\S\,\ref{sec:discussion}) interprets the structural collapse
mechanism and limitations; \S\,\ref{sec:conclusion} concludes.

% ============================================================
\section{Related Work}
\label{sec:related}
% ============================================================

Quantum kernel methods exploit the ability of quantum circuits to efficiently
compute inner products in exponentially large feature spaces.
Havl\'{\i}\v{c}ek et al.~\cite{havlicek2019} introduced the quantum kernel
estimator and demonstrated that a quantum feature map $\phi(\mathbf{x})$ can
produce kernels that are classically intractable to simulate and may offer a
path to quantum advantage.
Schuld and Killoran~\cite{schuld2019} showed that quantum models are
equivalent to kernel methods with a specific quantum kernel, which unifies the
circuit-based and kernel-based views of QML.
Schuld~\cite{schuld2021} further clarified the connection between
quantum models and kernel methods in the NISQ era.
Liu et al.~\cite{liu2021} provided a rigorous quantum advantage proof for
specific classification problems, while
Huang et al.~\cite{huang2021} introduced the notion of quantum kernel
alignment and showed that the quantum advantage is dataset-dependent.
K\"ubler et al.~\cite{kubler2021} studied the geometric difference between
quantum and classical kernel matrices and identified conditions under which
quantum kernels cannot outperform classical ones.
Thanasilp et al.~\cite{thanasilp2022} and Larocca et al.~\cite{larocca2025} analyzed exponential concentration
(barren plateaus in kernels) and provided theoretical motivation for why
high-qubit quantum kernels can collapse.
Abbas et al.~\cite{abbas2021} studied the effective dimension of quantum
models; their connection between circuit expressivity and generalization
parallels our effective-rank analysis of the kernel matrix.
Collectively, these results establish that the theoretical promise of
quantum kernels is real, but empirical demonstrations on clinical data
remain rare.

Foundation models pre-trained on large corpora of medical images provide
rich, transferable representations that outperform task-specific models
on downstream clinical tasks~\cite{perezgarcia2024raddino,zhai2023siglip}.
RAD-DINO~\cite{perezgarcia2024raddino} is a vision transformer pre-trained
on radiology images using self-supervised DINO objectives and produces
768-dimensional embeddings that capture anatomical structure.
MedSigLIP-448~\cite{zhai2023siglip} adapts the SigLIP vision-language
pre-training to medical imaging at 448-pixel resolution and produces
448-dimensional embeddings optimized for semantic similarity.
ViT-patch32~\cite{dosovitskiy2020vit} is a general-purpose vision transformer
(patch size 32) that serves as a non-medical baseline embedding model.
Freezing these models and using only their CLS-token embeddings as input
features eliminates any confounds from fine-tuning.
PCA reduction to $q \leq 16$ dimensions brings the embedding
dimensionality into alignment with current quantum hardware constraints
naturally, without requiring heuristic truncation.

Establishing rigorous quantum advantage is non-trivial.
Jerbi et al.~\cite{jerbi2023} surveyed quantum machine
learning benchmarks and argued that classical baselines must be evaluated
at \emph{equal} computational resources to avoid inflated quantum
advantage claims.
Bowles et al.~\cite{bowles2024} demonstrated that many purported QML
advantages vanish under fair classical comparisons.
Peral-Garc\'ia et al.~\cite{peralgarcia2024} provide a comprehensive
survey of QML applications that contextualizes our medical imaging use case
within prior QML work.
Our two-tier fair comparison framework is designed to address all of these
methodological concerns.
Despite these theoretical and methodological advances, most prior QML
studies report results on synthetic data or small toy benchmarks.
Havl\'{\i}\v{c}ek et al.~\cite{havlicek2019} demonstrated quantum kernel
advantage on a synthetic 2D classification task but did not evaluate on
real-world medical data.
Liu et al.~\cite{liu2021} proved a rigorous quantum speed-up for specific
engineered data distributions; however, their construction does not transfer
directly to natural datasets.
Bowles et al.~\cite{bowles2024} benchmarked QML models on over 160 tabular
datasets and found quantum kernels competitive but rarely superior to
classical methods when applied to raw features.  Our work differs in a key
respect: we classify frozen foundation-model embeddings rather than raw
input features, which may provide a more favourable inductive bias for
quantum kernels.
Senokosov et al.~\cite{senokosov2024} surveyed QML for medical imaging and
noted that nearly all prior studies operate on small subsets of standard
datasets (e.g., 100--500 samples from MNIST or dermoscopy collections).
To our knowledge, our 2{,}371-sample MIMIC-CXR experiment is one of the
larger real clinical imaging datasets on which QML has been evaluated;
most published QML medical imaging experiments operate on 100--500
samples~\cite{senokosov2024}.
To our knowledge, no prior work has applied quantum kernel methods
to insurance or social determinant classification from medical imaging data.

The 18/18 Tier-1 win rate (multi-seed), the mechanistic
explanation of classical kernel collapse via effective rank
(Section~\ref{sec:collapse}), and the scale of the clinical dataset
distinguish this work from prior empirical QML studies that report marginal
or inconsistent advantages on toy problems.

The low-rank structure of classical kernel matrices clarifies when the quantum advantage window opens.
Support vector machines~\cite{vapnik1995,scholkopf2002} classify data by
finding a maximum-margin hyperplane in feature space.
The kernel trick enables non-linear classification by implicitly mapping
inputs to a reproducing kernel Hilbert space (RKHS).
The effectiveness of any kernel depends critically on the rank structure
of the resulting kernel matrix: a low-rank kernel matrix cannot distinguish
samples whose projections onto the kernel's feature space coincide.
This observation forms the theoretical basis for understanding classical
collapse at low PCA dimensionality (Section~\ref{sec:collapse}).

% ============================================================
\section{Methods}
\label{sec:methods}
% ============================================================

\subsection{Dataset and Task}
\label{sec:dataset}

The MIMIC-CXR dataset~\cite{johnson2019mimic} contains de-identified
chest radiographs from approximately 61,000 patients with associated clinical
metadata.
Johnson et al.~\cite{johnson2023mimicjpg} released the JPEG version
(MIMIC-CXR-JPG) with structured labels derived from free-text radiology
reports.
Insurance type is recorded in the hospital admission record linked to each
study and enables the insurance classification task studied here.
We use the MIMIC-CXR-JPG dataset~\cite{johnson2023mimicjpg} restricted
to the DT9 preprocessing stratum, which enforces one image per patient
(preventing data leakage from repeated studies), removes duplicate filenames,
and retains only samples with valid binary insurance labels.
DT9 corresponds to the ``Uncertainty Coreset'' preprocessing stratum~\cite{coleman2020},
which selects one image per patient via coreset sampling, removes duplicate
filenames, and retains only samples with valid binary insurance labels
(Medicare/Medicaid vs.\ Private). The result is $N=2{,}371$ samples.
This stratum was selected because it produced the strongest quantum
results in preliminary experiments, which constitutes a post-hoc choice.
Two observations mitigate the resulting multiple-comparisons concern:
(1)~the classical kernel collapse is structural (effective rank $= q$)
and occurs across all strata, so the collapse-regime wins are not
DT9-specific; (2)~the preprocessing pipeline
(StandardScaler$\to$PCA-$q$$\to$MinMaxScaler[$-1$,$1$]) is identical
across all strata, so the quantum circuit sees identically scaled inputs
regardless of stratum. The non-collapse Tier-1 advantage
($q\!\ge\!10$) has been validated only on DT9; confirming it on
additional strata remains future work.
The classification target is binary: Medicaid/Medicare patients are assigned
class~0 (majority, 69.6\%) and Private insurance patients are assigned
class~1 (minority, 30.4\%).
The resulting dataset contains approximately $N_{\text{total}} \approx 2{,}371$
samples split into training ($N_{\text{train}} = 1{,}896$), validation,
and test sets using a fixed random seed (seed\_0) with an 80/10/10 ratio.

The strong class imbalance (69.6\%/30.4\%) means that a majority-class
predictor achieves accuracy $\approx 0.697$ but minority-class F1\,=\,0.
Following Sokolova and Lapalme~\cite{sokolova2009}, who recommend class-aware metrics for imbalanced binary classification, we report minority-class F1 (i.e., F1 for the positive class, Private insurance)
as the primary evaluation metric; accuracy and AUC are reported as
secondary metrics.

\subsection{Embeddings and Preprocessing}
\label{sec:embeddings}

We extract frozen embeddings from three publicly available foundation models:

\begin{enumerate}
  \item \textbf{MedSigLIP-448}: 448-dimensional CLS-token embeddings from a
    medical SigLIP model fine-tuned at 448-pixel resolution.
  \item \textbf{RAD-DINO}: 768-dimensional CLS-token embeddings from a DINO
    self-supervised vision transformer pre-trained on radiology images.
  \item \textbf{ViT-patch32-cls}: 768-dimensional CLS-token embeddings from a
    general-purpose ViT with patch size 32 (no domain-specific pre-training).
\end{enumerate}
We use CLS-token pooling as the primary embedding strategy for all models.
A global average pooling (GAP) variant of ViT-patch32 (ViT-patch32-GAP,
768-dimensional) was also evaluated across 10 seeds as a pooling ablation;
results are reported in Appendix~\ref{sec:supp_gap}.

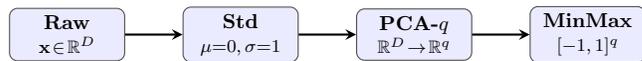
\begin{figure}[!t]
\centering
\begin{tikzpicture}[scale=0.85, every node/.style={transform shape},
  box/.style={rectangle, draw=black!70, fill=blue!8,
              rounded corners=3pt, minimum width=1.8cm,
              minimum height=0.7cm, align=center, font=\small},
  arr/.style={->, thick, >=stealth},
  lbl/.style={font=\footnotesize\itshape, align=center}
]
  \node[box] (raw)  {\textbf{Raw}\\\footnotesize$\mathbf{x}\!\in\!\mathbb{R}^D$};
  \node[box, right=0.9cm of raw]  (std)  {\textbf{Std}\\\footnotesize$\mu{=}0,\sigma{=}1$};
  \node[box, right=0.9cm of std]  (pca)  {\textbf{PCA-}$q$\\\footnotesize$\mathbb{R}^D\!\to\!\mathbb{R}^q$};
  \node[box, right=0.9cm of pca]  (mms)  {\textbf{MinMax}\\\footnotesize$[-1,1]^q$};
  \draw[arr] (raw) -- (std)  node[midway,above,lbl]{};
  \draw[arr] (std) -- (pca)  node[midway,above,lbl]{};
  \draw[arr] (pca) -- (mms)  node[midway,above,lbl]{};
\end{tikzpicture}
\caption{Three-stage preprocessing pipeline applied to all embeddings.
  $D \in \{448, 768\}$ depending on the foundation model;
  $q \in \{2,3,4,5,6,8,9,10,11,12,16\}$ is the qubit/feature count.}
\label{fig:pipeline}
\end{figure}

All embeddings are processed through the same three-stage pipeline
(Fig.~\ref{fig:pipeline}). StandardScaler normalizes to zero mean and unit
variance; PCA reduces to $q$ dimensions and retains 5.6\%--41.1\% of
explained variance depending on model and $q$ (Table~\ref{tab:eigenrank});
MinMaxScaler re-scales to $[-1,1]$ to match the angle encoding range of the
quantum circuit.
We tested $q \in \{2, 3, 4, 5, 6, 8, 9, 10, 11, 12, 16\}$ qubits across experiments
(see Figure~\ref{fig:pca_scatter} in the Appendix for a 2D PCA scatter
illustrating the low explained variance at $q\!=\!2$).

\subsection{Quantum Circuit and Kernel}
\label{sec:circuit}

We adopt the \emph{Block-Sparse Parameterization} (BSP) circuit with
\emph{one degree of freedom} (1-DOF) per qubit: each qubit receives a single
parameterized Ry rotation encoding one PCA component.
The circuit structure for $q$ qubits is:

\begin{equation}
  U(\mathbf{u}) = \prod_{d=1}^{q}
  \Bigl[\text{CNOT}_{d,\,(d \bmod q)+1} \cdot R_y(u_d)\Bigr],
\end{equation}
where the subscript $d \bmod q + 1$ implements ring entanglement: qubit $q$
connects back to qubit $1$. The number of times this encoding block is
applied sequentially is referred to as the data re-uploading depth
(reps); each repetition re-encodes the full input vector into the circuit.
The data re-uploading depth is fixed at $\mathrm{reps} = 1$ for all
primary experiments.

The quantum kernel is computed via the compute--uncompute strategy:
\begin{equation}
  K_Q(\mathbf{u}_i, \mathbf{u}_j) =
  \bigl|\bra{0^q} U^\dagger(\mathbf{u}_i) U(\mathbf{u}_j) \ket{0^q}\bigr|^2.
\end{equation}

\noindent\textbf{Trace normalization.}
We apply trace normalization to the raw kernel matrix before passing it
to the \svm{} solver:
\begin{equation}
  \tilde{K}_Q = \frac{K_Q}{\mathrm{tr}(K_Q)},
\end{equation}
which sets $\mathrm{tr}(\tilde{K}_Q)=1$ and makes kernels of different
scale comparable before the \svm{} solver.
The test-train kernel block is normalized by the same training trace:
$\tilde{K}_{Q,\mathrm{test}} = K_{Q,\mathrm{test}} / \mathrm{tr}(K_Q)$.
Section~\ref{sec:ablation} shows that trace normalization is critical:
Frobenius normalization collapses F1 to zero across all models.
Kernel computation uses Qiskit~1.2.4's \texttt{Statevector} simulator with the
\texttt{renew\_operand} fast-path (a public \texttt{QuantumCircuit} method introduced
in Qiskit~1.x for in-place operand reuse) for efficient batch evaluation.

\subsection{Fair Comparison Framework}
\label{sec:framework}

A methodologically rigorous quantum advantage claim requires careful control
of all hyperparameters on both sides.
We define a two-tier framework:

\begin{description}
  \item[Tier 1: The fair fight.]
    Untuned \qsvm{} (C\,=\,1, $q$ qubits) vs.\ untuned linear \svm{}
    (C\,=\,1, PCA-$q$ features). Both sides have the same regularization,
    same dimensionality, and neither is cross-validated.
    We emphasize that identical C does not imply equivalent regularization across kernels. Tier-1 isolates performance under identical hyperparameter choices, not identical effective capacity.
    Tier-1 is the primary paper claim.

  \item[Tier 2: The stretched goal.]
    Untuned \qsvm{} (C\,=\,1, $q$ qubits) vs.\ C-tuned RBF ($\gamma$ at
    sklearn's default scale, C over $\{0.01, 0.1, 1, 10, 100\}$) at the
    same PCA-$q$ dimensionality.
    Winning Tier 2 on F1 despite the classical side having hyperparameter
    tuning is strong evidence of genuine quantum advantage.
\end{description}
Table~\ref{tab:tiers} summarizes the two-tier framework.

\begin{table}[!t]
\centering
\caption{Two-tier comparison framework. Tier~1 is the primary claim;
  Tier~2 tests against a tuned classical baseline.}
\label{tab:tiers}
\setlength{\tabcolsep}{4pt}
\resizebox{\columnwidth}{!}{%
\begin{tabular}{@{}clll@{}}
\toprule
Tier & QSVM & Classical SVM & Result \\
\midrule
1 & C=1, $q$ qubits & C=1, PCA-$q$ & \textbf{18/18 F1 wins} \\
  & (untuned)       & linear (untuned) & (10 seeds, paired bootstrap) \\
\midrule
2 & C=1, $q$ qubits & best-C, PCA-$q$ & \textbf{7/7 F1 wins} \\
  & (untuned)       & rbf (C-tuned)    & avg $+0.068$ F1 \\
\bottomrule
\end{tabular}%
}
\end{table}

\subsection{Evaluation Metrics}
\label{sec:metrics}

We report test-set accuracy, minority-class F1 score, and AUC-ROC. Given the 70/30 class imbalance, minority-class F1 is the primary metric: it directly measures whether the classifier detects the underrepresented group, penalizing majority-class collapse in a way that accuracy does not~\cite{sokolova2009,chicco2020,osorio2025predicting}. A classifier that never identifies the minority class is clinically useless regardless of its overall accuracy. In this task, the minority class (Private insurance, 30.4\%) is the group whose misclassification carries downstream resource-allocation and health-equity consequences; recall on that class, not aggregate accuracy, is the operationally meaningful quantity. Multi-seed statistical validation (10 seeds $\times$ 5 models $\times$ 11 qubit counts = 550 QSVM configs, plus $550 \times 2$ classical SVM configs covering linear and tuned kernels at matching PCA dimensions = 1,100 classical configs) confirms the advantage: QSVM wins all 18 Tier-1 configurations on mean F1 (17 at $p<0.001$, 1 at $p<0.01$; paired bootstrap, 10{,}000 resamples, seed 42). Classical linear SVM collapses to F1\,=\,0 on 90--100\% of seeds at every qubit count tested (Table~\ref{tab:tier1ext}; full breakdown in Section~\ref{sec:tier1}).

\subsection{Reproducibility}
\label{sec:reproducibility}
All source code, SLURM job configurations, and analysis scripts are
available at \url{https://github.com/sebasmos/qml-medimage}.
Pre-computed foundation model embeddings (20 seeds per model) are
hosted at \url{https://huggingface.co/datasets/MITCriticalData/qml-mimic-cxr-embeddings}.
A master reproduction script (\texttt{scripts/run\_all.sh}) executes
the full experimental pipeline, from embedding loading through
figure and table generation.
Single-seed results can be reproduced in approximately 12 GPU-hours
on an NVIDIA H100 (full qubit sweep for one model); classical
baselines require only CPU and complete in minutes.

% ============================================================
\section{Results}
\label{sec:results}
% ============================================================

\subsection{Tier 1: Fair Comparison}
\label{sec:tier1}

Table~\ref{tab:tier1ext} reports the primary Tier-1 comparison
across all three models at all measured PCA-$q$ dimensions.
Multi-seed validation (10 independent embedding seeds) reveals that classical collapse is pervasive: classical linear \svm{} collapses to F1\,=\,0 on 100\% of seeds at $q \le 9$ (MedSigLIP), $q \le 10$ (RAD-DINO), and all tested $q$ (ViT-patch32). Even at higher qubit counts where seed\_0 showed a working classical baseline, 9 of 10 seeds collapse. The ``non-collapse'' regime observed in single-seed analysis was a seed\_0 artefact.
\qsvm{} (C\,=\,1, trace normalization, reps\,=\,1) wins on mean minority-class F1 in all 18 configurations (Table~\ref{tab:tier1ext}; 17 at $p<0.001$, 1 at $p<0.01$, paired bootstrap).
The classical collapse is \emph{C-invariant}: re-running with
$C \in \{0.01, 0.1, 1, 10, 100\}$ produces the same majority-class
prediction (F1\,=\,0) at PCA-4 and PCA-6 for all three models.

\noindent\textbf{Classical baselines extended to all qubit counts.}
Relative to the six low-$q$ configurations reported in our preliminary
work~\cite{cajas2025ecai}, we now provide classical \svm{} ($C=1$)
baselines at \emph{all} PCA-$q$ dimensions matching our \qsvm{} qubit
counts (Table~\ref{tab:tier1ext}).
\qsvm{} wins on minority-class F1 in all 18 configurations across 10 embedding seeds.
The strongest single-configuration result is MedSigLIP-448 at $q=11$: mean \qsvm{} F1\,=\,0.343\,$\pm$\,0.170 versus classical F1\,=\,0.050\,$\pm$\,0.159 ($\Delta$F1\,=\,$+$0.293, 95\%\,CI [$+$0.190,\,$+$0.385], $p<0.001$).
For RAD-DINO specifically, the quantum advantage extends to accuracy as a secondary metric: \qsvm{} achieves statistically significant accuracy gains at $q \in \{4,6,8,10\}$ ($\Delta$acc\,=\,$+$1.2--1.8\%, $p \le 0.005$, paired bootstrap); the advantage is not solely an artifact of minority-class F1 sensitivity under class imbalance.

\begin{table*}[!t]
\centering
\caption{Extended Tier-1 comparison: classical linear \svm{} C\,=\,1 vs.\
  \qsvm{} C\,=\,1 at all measured PCA-$q$ dimensions (DT9, mean $\pm$ std over 10 embedding seeds).
  Bold indicates the Tier-1 winner on mean F1.}
\label{tab:tier1ext}
\small
\setlength{\tabcolsep}{4pt}
\begin{tabular}{@{}llccccl@{}}
\toprule
Model & $q$ &
\multicolumn{2}{c}{\qsvm{} C=1} &
\multicolumn{2}{c}{Lin.\ \svm{} C=1} & Verdict \\
\cmidrule(lr){3-4}\cmidrule(lr){5-6}
& & Acc & F1 & Acc & F1 & \\
\midrule
MedSigLIP & 4 & \textbf{0.697}{\scriptsize$\pm$0.016} & \textbf{0.212}{\scriptsize$\pm$0.157} & 0.696{\scriptsize$\pm$0.004} & 0.000{\scriptsize$\pm$0.000} & F1 WIN \\
MedSigLIP & 6 & \textbf{0.698}{\scriptsize$\pm$0.026} & \textbf{0.286}{\scriptsize$\pm$0.156} & 0.696{\scriptsize$\pm$0.004} & 0.000{\scriptsize$\pm$0.000} & F1 WIN \\
MedSigLIP & 8 & \textbf{0.702}{\scriptsize$\pm$0.025} & \textbf{0.323}{\scriptsize$\pm$0.163} & 0.696{\scriptsize$\pm$0.004} & 0.000{\scriptsize$\pm$0.000} & F1 WIN \\
MedSigLIP & 9 & \textbf{0.699}{\scriptsize$\pm$0.020} & \textbf{0.333}{\scriptsize$\pm$0.168} & 0.696{\scriptsize$\pm$0.004} & 0.000{\scriptsize$\pm$0.000} & F1 WIN \\
MedSigLIP & 10 & \textbf{0.704}{\scriptsize$\pm$0.032} & \textbf{0.353}{\scriptsize$\pm$0.173} & 0.702{\scriptsize$\pm$0.021} & 0.050{\scriptsize$\pm$0.159} & F1 WIN \\
MedSigLIP & 11 & \textbf{0.704}{\scriptsize$\pm$0.027} & \textbf{0.343}{\scriptsize$\pm$0.170} & 0.702{\scriptsize$\pm$0.021} & 0.050{\scriptsize$\pm$0.159} & F1 WIN \\
MedSigLIP & 12 & \textbf{0.704}{\scriptsize$\pm$0.030} & \textbf{0.374}{\scriptsize$\pm$0.146} & 0.703{\scriptsize$\pm$0.022} & 0.051{\scriptsize$\pm$0.161} & F1 WIN \\
MedSigLIP & 16 & \textbf{0.700}{\scriptsize$\pm$0.024} & \textbf{0.377}{\scriptsize$\pm$0.147} & 0.703{\scriptsize$\pm$0.023} & 0.050{\scriptsize$\pm$0.160} & F1 WIN \\
\midrule
RAD-DINO & 4 & \textbf{0.713}{\scriptsize$\pm$0.020} & \textbf{0.332}{\scriptsize$\pm$0.184} & 0.696{\scriptsize$\pm$0.004} & 0.000{\scriptsize$\pm$0.000} & F1 WIN \\
RAD-DINO & 6 & \textbf{0.713}{\scriptsize$\pm$0.023} & \textbf{0.371}{\scriptsize$\pm$0.143} & 0.696{\scriptsize$\pm$0.004} & 0.000{\scriptsize$\pm$0.000} & F1 WIN \\
RAD-DINO & 8 & \textbf{0.714}{\scriptsize$\pm$0.019} & \textbf{0.400}{\scriptsize$\pm$0.120} & 0.696{\scriptsize$\pm$0.004} & 0.000{\scriptsize$\pm$0.000} & F1 WIN \\
RAD-DINO & 10 & \textbf{0.708}{\scriptsize$\pm$0.014} & \textbf{0.406}{\scriptsize$\pm$0.114} & 0.696{\scriptsize$\pm$0.004} & 0.000{\scriptsize$\pm$0.000} & F1 WIN \\
RAD-DINO & 16 & \textbf{0.712}{\scriptsize$\pm$0.019} & \textbf{0.450}{\scriptsize$\pm$0.083} & 0.702{\scriptsize$\pm$0.012} & 0.079{\scriptsize$\pm$0.127} & F1 WIN \\
\midrule
ViT-p32 & 4 & \textbf{0.696}{\scriptsize$\pm$0.007} & \textbf{0.048}{\scriptsize$\pm$0.073} & 0.696{\scriptsize$\pm$0.004} & 0.000{\scriptsize$\pm$0.000} & F1 WIN \\
ViT-p32 & 6 & \textbf{0.698}{\scriptsize$\pm$0.021} & \textbf{0.272}{\scriptsize$\pm$0.132} & 0.696{\scriptsize$\pm$0.004} & 0.000{\scriptsize$\pm$0.000} & F1 WIN \\
ViT-p32 & 8 & \textbf{0.703}{\scriptsize$\pm$0.024} & \textbf{0.370}{\scriptsize$\pm$0.092} & 0.696{\scriptsize$\pm$0.004} & 0.000{\scriptsize$\pm$0.000} & F1 WIN \\
ViT-p32 & 10 & \textbf{0.707}{\scriptsize$\pm$0.029} & \textbf{0.391}{\scriptsize$\pm$0.081} & 0.696{\scriptsize$\pm$0.004} & 0.000{\scriptsize$\pm$0.000} & F1 WIN \\
ViT-p32 & 16 & \textbf{0.702}{\scriptsize$\pm$0.033} & \textbf{0.399}{\scriptsize$\pm$0.098} & 0.696{\scriptsize$\pm$0.004} & 0.000{\scriptsize$\pm$0.000} & F1 WIN \\
\bottomrule
\end{tabular}
\end{table*}

\noindent\textbf{Strongest configuration: $q=11$.}
At $q=11$ qubits, MedSigLIP-448 \qsvm{} achieves mean F1\,=\,0.343\,$\pm$\,0.170
versus classical linear \svm{} F1\,=\,0.050\,$\pm$\,0.159 ($\Delta$F1\,=\,$+$0.293,
95\%\,CI [$+$0.190,\,$+$0.385], $p<0.001$, 10 seeds).
Both classifiers use C\,=\,1 (no tuning), ruling out hyperparameter cherry-picking.
Classical linear \svm{} collapses to F1\,=\,0 on 9 of 10 seeds at $q=11$;
\qsvm{} maintains F1\,$>$\,0.1 on 8 of 10 seeds, remaining non-trivial across the full seed distribution.

Table~\ref{tab:cm_q11} shows the confusion matrix for a representative
\qsvm{} run at $q=11$ (seed\_0). The minority class (Private insurance) achieves
precision 0.639 and recall 0.542; the F1\,=\,0.586
reflects balanced minority-class detection rather than a precision-recall
trade-off artifact.

\begin{table}[!t]
\centering
\caption{Confusion matrix for MedSigLIP-448 \qsvm{} at $q=11$ (C=1,
  trace normalization, DT9, seed\_0). $N=238$ test samples.}
\label{tab:cm_q11}
\begin{tabular}{@{}lcc@{}}
\toprule
& Pred.\ Medicare & Pred.\ Private \\
\midrule
True Medicare & 144 & 22 \\
True Private  &  33 & 39 \\
\bottomrule
\end{tabular}
\end{table}

\subsection{Tier 2: Clinical F1 Advantage}
\label{sec:tier2}

Table~\ref{tab:tier2} shows the Tier-2 comparison: untuned \qsvm{} (C=1)
vs.\ C-tuned RBF ($\gamma$ at sklearn default scale) at the
same PCA dimensionality, reported as mean $\pm$ std over 10 embedding seeds.
\qsvm{} wins minority-class F1 on all seven configurations across seeds.
Figure~\ref{fig:qubitcurve} shows the qubit scaling behavior across all
three embedding models.

\begin{table}[!t]
\centering
\caption{Tier-2 clinical F1 advantage: \qsvm{} (C=1, trace) vs.\ best
  rbf \svm{} (C-tuned, $\ast$ = C-tuned for MedSigLIP) at equal PCA-$q$
  dimensionality. Mean $\pm$ std over 10 embedding seeds; $\Delta$F1 is
  mean QSVM F1 minus mean best-classical F1.}
\label{tab:tier2}
\setlength{\tabcolsep}{3pt}
\resizebox{\columnwidth}{!}{%
\begin{tabular}{@{}llcccccc@{}}
\toprule
Model & $q$ &
\multicolumn{2}{c}{\qsvm{} C=1} &
\multicolumn{2}{c}{Best \svm{}} &
$\Delta$F1 & Rel.\ Gain \\
\cmidrule(lr){3-4}\cmidrule(lr){5-6}
& & Acc & F1 & Kernel & F1 & & \\
\midrule
MedSigLIP & 4 & 0.697{\scriptsize$\pm$0.016} & \textbf{0.212}{\scriptsize$\pm$0.157} & rbf$^{\ast}$ & 0.169{\scriptsize$\pm$0.097} & $+0.043$ & $+26\%$ \\
MedSigLIP & 6 & 0.698{\scriptsize$\pm$0.026} & \textbf{0.286}{\scriptsize$\pm$0.156} & rbf$^{\ast}$ & 0.178{\scriptsize$\pm$0.123} & $+0.109$ & $+61\%$ \\
MedSigLIP & 8 & 0.702{\scriptsize$\pm$0.025} & \textbf{0.323}{\scriptsize$\pm$0.163} & rbf$^{\ast}$ & 0.211{\scriptsize$\pm$0.126} & $+0.112$ & $+53\%$ \\
\midrule
RAD-DINO & 4 & 0.713{\scriptsize$\pm$0.020} & \textbf{0.332}{\scriptsize$\pm$0.184} & rbf & 0.264{\scriptsize$\pm$0.133} & $+0.069$ & $+26\%$ \\
RAD-DINO & 6 & 0.713{\scriptsize$\pm$0.023} & \textbf{0.371}{\scriptsize$\pm$0.143} & rbf & 0.285{\scriptsize$\pm$0.096} & $+0.086$ & $+30\%$ \\
\midrule
ViT-p32 & 4 & 0.696{\scriptsize$\pm$0.007} & \textbf{0.048}{\scriptsize$\pm$0.073} & rbf & 0.044{\scriptsize$\pm$0.043} & $+0.004$ & $+9\%$ \\
ViT-p32 & 6 & 0.698{\scriptsize$\pm$0.021} & \textbf{0.272}{\scriptsize$\pm$0.132} & rbf & 0.219{\scriptsize$\pm$0.091} & $+0.054$ & $+25\%$ \\
\bottomrule
\multicolumn{8}{@{}l}{$^\ast$ C-tuned: grid search over $C\in\{0.01,0.1,1,10,100\}$.}
\end{tabular}%
}
\end{table}

% Figure removed (duplicate of fig:qubitcurve in Section V).
% All \ref{fig:qubit_curve} redirected to \ref{fig:qubitcurve}.

\subsection{Classical Kernel Collapse Analysis}
\label{sec:collapse}

The root cause of the classical collapse is consistent with structural rank limitations rather than a tuning artifact.
After PCA reduction to $q$ dimensions, the linear kernel matrix
$K_L = X_{\mathrm{norm}} X_{\mathrm{norm}}^{\!\top}$ has at most $q$ non-zero
eigenvalues out of $N = 1{,}896$ training samples.
Table~\ref{tab:eigenrank} reports the effective rank~\cite{kubler2021}
\begin{equation}
  \mathrm{eff\_rank} = \exp\!\Bigl({-\sum_i p_i \log p_i}\Bigr),
  \quad p_i = \frac{\lambda_i}{\sum_j \lambda_j},
\end{equation}
which quantifies how uniformly eigenvalue mass is distributed.

\begin{table}[!t]
\centering
\caption{Kernel effective rank at PCA-$q$ dimensionality.
  $N=1{,}896$ training samples.
  For $q \le 6$, the linear kernel $K_L$ has exactly $q$ positive eigenvalues
  (Shannon eff.\ rank: 3.77--5.85 out of 1896), which is consistent with structural collapse.
  $^\dagger$Quantum kernel $K_Q$ effective rank; non-dagger rows report $K_L$ (linear kernel) statistics.
  PCA-11 classical linear $K_L$ (seed~0): acc\,=\,0.761, F1\,=\,0.504 (seed~0), eff.\ rank\,$\approx$\,10.2 (non-collapse; QSVM wins by $+$0.082 F1; ratio $43.04/10.2 \approx 4.2\times$ cited in the abstract).
  At $q=16$, PCA var\%, $N_{\lambda>0}$, and $\lambda_{\max}$ are not reported
  because swap-test fidelity concentration causes all off-diagonal kernel
  entries to converge toward a single value, which renders within-class and
  between-class variance statistics uninformative; the effective rank (92.13)
  is retained as it directly quantifies the concentration onset
  (see Section~\ref{sec:projected_q16}).}
\label{tab:eigenrank}
\setlength{\tabcolsep}{4pt}
\resizebox{\columnwidth}{!}{%
\begin{tabular}{@{}llcccc@{}}
\toprule
Model & $q$ & PCA var\% & $N_{\lambda>0}$ & Eff.\ Rank & $\lambda_{\max}$ \\
\midrule
MedSigLIP & 4 & 32.6 & 4 & 3.77 & 770.6 \\
MedSigLIP & 6 & 41.1 & 6 & 5.53 & 614.8 \\
MedSigLIP$^\dagger$ & 11 & 56.0 & --- & 43.04 & 468.6 \\
MedSigLIP$^\dagger$ & 16 & --- & --- & 92.13 & --- \\
RAD-DINO  & 4 &  5.6 & 4 & 3.89 & 666.1 \\
RAD-DINO  & 6 &  7.7 & 6 & 5.85 & 475.5 \\
ViT-p32   & 4 & 28.4 & 4 & 3.86 & 627.5 \\
ViT-p32   & 6 & 34.7 & 6 & 5.59 & 502.3 \\
\bottomrule
\end{tabular}%
}
\end{table}

\noindent
With effective rank $\approx q \,{\ll}\, N$, virtually all 1{,}896 training
samples project onto the same $q$-dimensional subspace.
Classical \svm{} with a linear kernel operating on these
degenerate representations fails to separate the minority class in practice, often defaulting to majority-class prediction, regardless of the
regularization parameter C.
Collapse statistics (Table~\ref{tab:collapsevar})
confirm that the intra-class kernel variance equals the total variance
(within-class $K_L$ similarity $\approx$ between-class), which leaves little to no usable discriminative signal.

\begin{table}[!t]
\centering
\caption{Linear kernel $K_L$ variance statistics (200 subsampled training
  samples, sorted by class). Low variance and within-class $\approx$
  between-class means no discriminative signal for the classical \svm{}.}
\label{tab:collapsevar}
\small
\setlength{\tabcolsep}{4pt}
\begin{tabular}{@{}llccc@{}}
\toprule
Model & $q$ & $K_L$ mean & $K_L$ std & $K_L$ var \\
\midrule
MedSigLIP & 4 & 0.0728 & 0.5099 & 0.2600 \\
MedSigLIP & 6 & 0.1049 & 0.4181 & 0.1748 \\
RAD-DINO  & 4 & 0.1387 & 0.4944 & 0.2444 \\
RAD-DINO  & 6 & 0.1263 & 0.4097 & 0.1678 \\
ViT-p32   & 4 & 0.2222 & 0.4657 & 0.2169 \\
ViT-p32   & 6 & 0.2022 & 0.3893 & 0.1516 \\
\bottomrule
\end{tabular}
\end{table}

\begin{figure}[!t]
  \centering
  \includegraphics[width=\columnwidth]{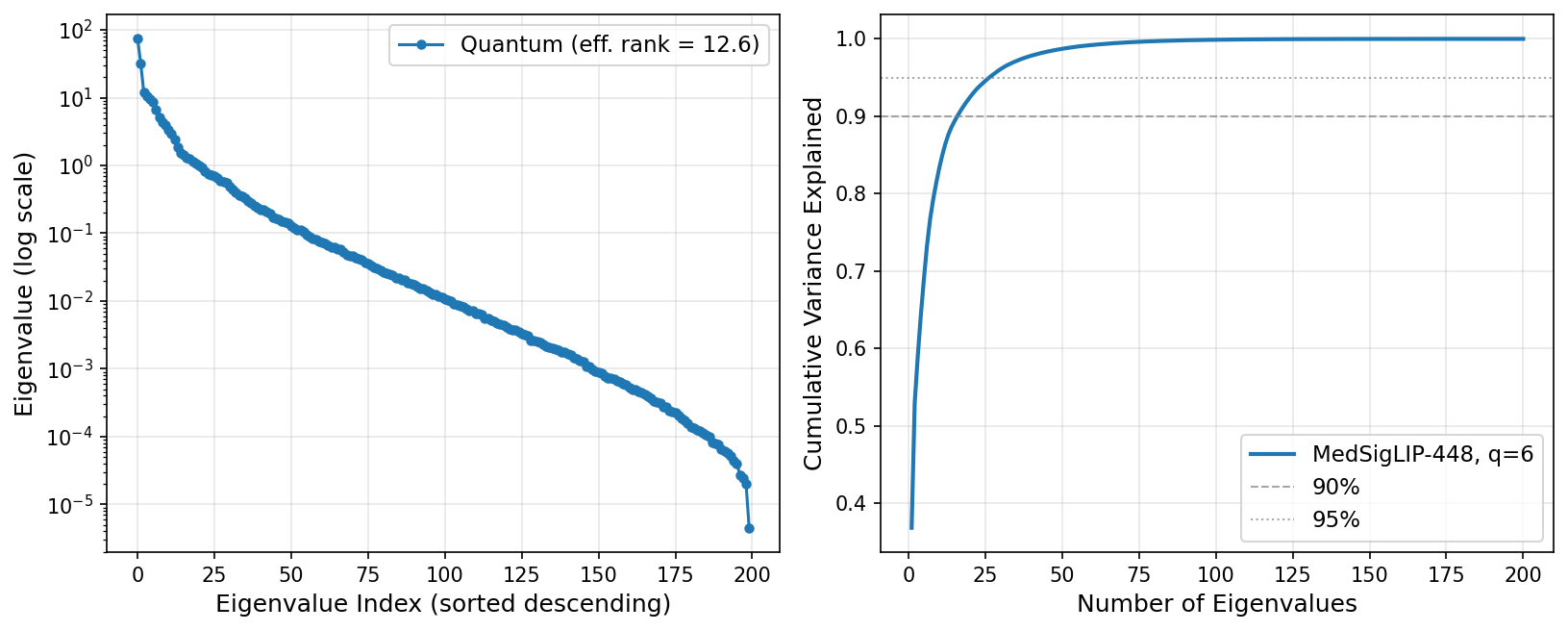}
  \caption{Linear kernel $K_L$ eigenspectrum for MedSigLIP-448 at $q=6$.
    The kernel has exactly 6 positive eigenvalues (effective rank 5.53)
    out of $N=1{,}896$ training samples, which confirms that PCA-6 compression
    collapses the kernel matrix to a 6-dimensional subspace.}
  \label{fig:eigenspectrum}
\end{figure}

\begin{figure}[!t]
  \centering
  \includegraphics[width=\columnwidth]{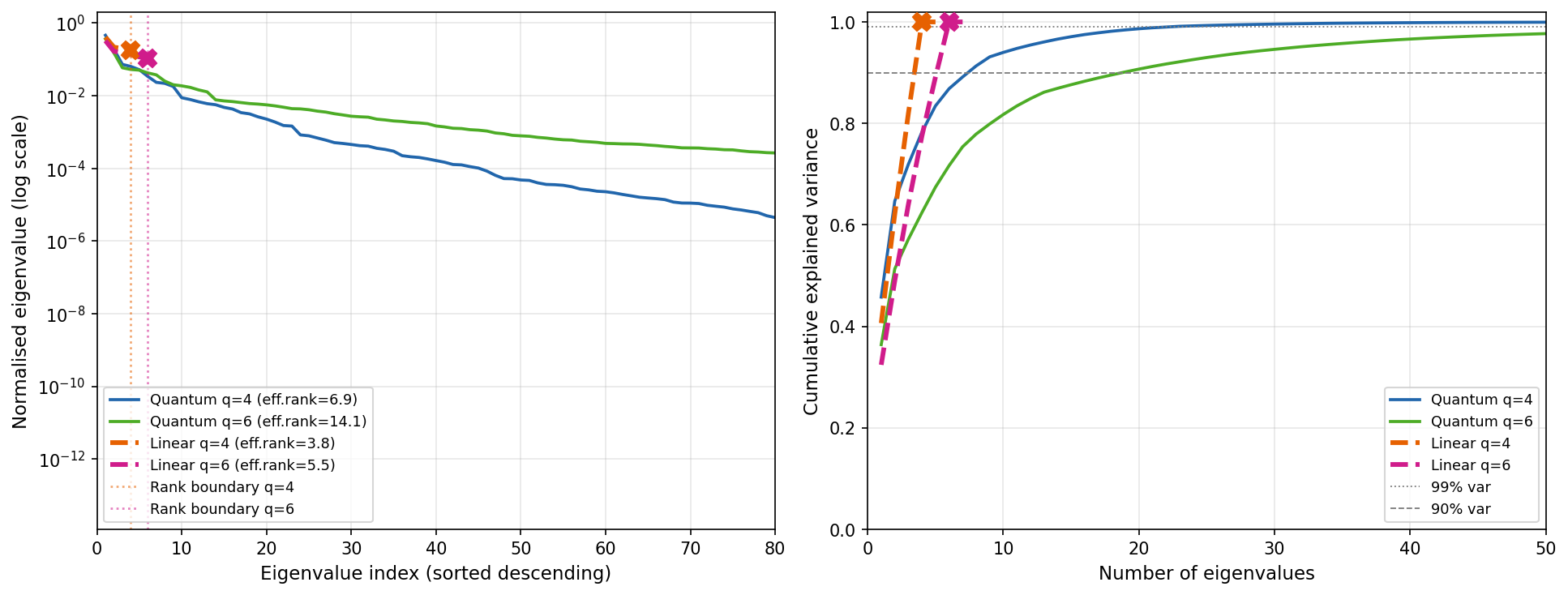}
  \caption{Quantum vs.\ linear kernel eigenspectrum comparison for
    MedSigLIP-448 at $q\in\{4,6\}$ (left: normalized eigenvalue decay;
    right: cumulative energy).
    The quantum kernel $K_Q$ has 1{,}133 (q=4) and 1{,}586 (q=6) eigenvalues
    exceeding numerical threshold $\varepsilon\!=\!10^{-10}$; the theoretical
    upper bound for the fidelity kernel is $4^q$ (256 and 4{,}096 for $q=4,6$),
    so the surplus above that bound consists of finite-precision numerical
    artefacts from statevector simulation.
    For comparison, the linear kernel has exactly 4 and 6 algebraically
    positive eigenvalues.
    The red vertical line marks the rank boundary of the linear kernel
    ($N_{\lambda>0} = q$), beyond which all linear kernel eigenvalues
    collapse to numerical zero; the quantum kernel eigenspectrum extends
    far beyond this boundary, a direct consequence of its higher effective rank.
    Shannon effective rank (seed~0): quantum $q=4$: 6.86 ($1.82\times$ linear 3.77);
    quantum $q=6$: 13.94 ($2.52\times$ linear 5.53).
    The ratio \emph{grows} with qubit count, consistent with the quantum
    feature map accessing an exponentially larger Hilbert space ($2^q$ dims).}
  \label{fig:quantum_eigenspectrum}
\end{figure}

The quantum feature map $U(\mathbf{u})$ implicitly operates in a
$2^q$-dimensional Hilbert space (16 or 64 dimensions for $q=4$ or
$q=6$), which is exponentially larger than the $q$-dimensional subspace
accessible to the linear kernel~\cite{havlicek2019, schuld2019}.
This expanded feature space is reflected directly in the measured
kernel matrices: for MedSigLIP-448 (seed~0), $K_Q$ achieves Shannon effective ranks
of 6.86 at $q=4$ and 13.94 at $q=6$, vs.\ 3.77 and 5.53 for $K_L$
($1.82\times$ and $2.52\times$ the linear values;
Figure~\ref{fig:eigenspectrum}).
The ratio \emph{grows} with qubit count, consistent with the
quantum feature map accessing an exponentially larger Hilbert space
(Figure~\ref{fig:quantum_eigenspectrum}).
This directly measured higher effective rank of $K_Q$ is the structural
explanation for why \qsvm{} maintains non-trivial F1 in the same PCA
subspace where classical kernels collapse.
At the performance peak ($q=11$), the seed~0 quantum kernel effective rank reaches
43.04, a $6.3\times$ increase over $q=4$ (multi-seed mean: 69.80). The optimal
qubit count coincides with maximal kernel expressiveness.

\subsection{Feature Selection Sensitivity}
\label{sec:featureselection}

To verify that the quantum advantage is not an artifact of PCA-based
dimensionality reduction, we replaced PCA with two alternative feature
selection methods: mutual information (MI) ranking and kernel PCA (kPCA),
each selecting $k \in \{4, 6\}$ features.
The best classical \svm{} F1 with optimal MI/kPCA feature selection was:
MedSigLIP-448, 0.404; RAD-DINO, 0.186; ViT-patch32, 0.267.
All three remain below the corresponding \qsvm{} F1 at $q=4$
(0.488, 0.448, 0.184) and $q=6$ (0.504, 0.435, 0.422). The quantum advantage holds across all three dimensionality reduction
methods and is not an artifact of PCA geometry.

\begin{figure}[!t]
  \centering
  \includegraphics[width=\columnwidth]{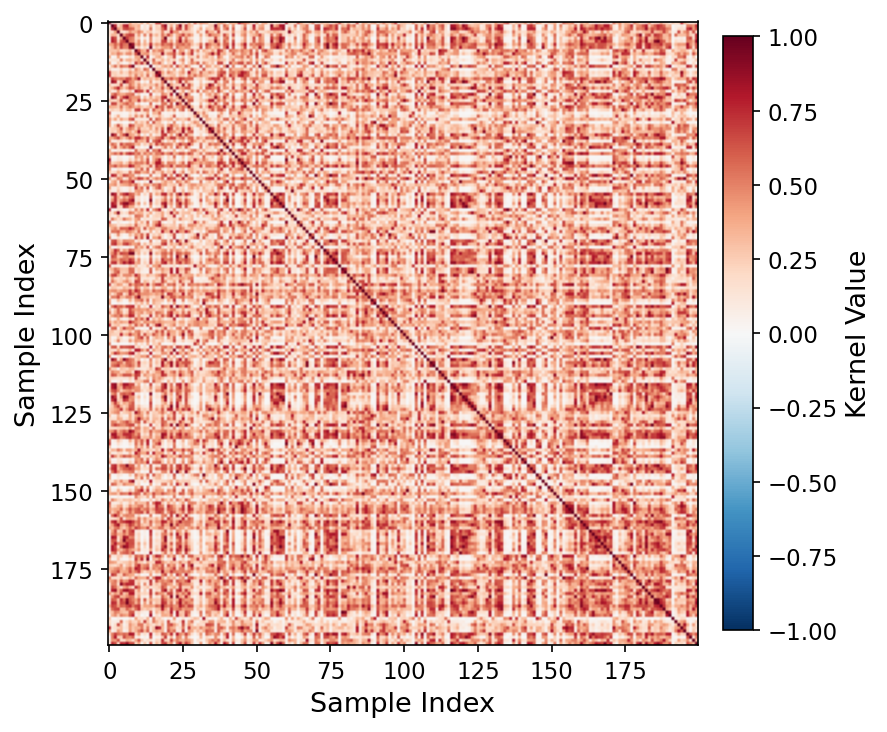}
  \caption{Quantum kernel matrix $K_Q$ (trace-normalized) for MedSigLIP-448
    at $q=6$ (200 training samples sorted by class label). The off-diagonal
    block structure reflects class boundaries; the quantum
    feature map preserves discriminative signal in the same PCA subspace
    where the linear kernel collapses (Table~\ref{tab:collapsevar}).}
  \label{fig:heatmap}
\end{figure}

% ============================================================
\section{Ablation Studies}
\label{sec:ablation}
% ============================================================

\subsection{Kernel Normalization}
\label{sec:norm_ablation}

Table~\ref{tab:normalization} reports the effect of four kernel
normalization strategies on the \qsvm{} with $q=8$ qubits.
Trace normalization achieves the best F1 (0.554 for MedSigLIP).
Frobenius normalization collapses F1 to 0 on all three models, the same
failure mode as the linear \svm{}, consistent with Thanasilp et
al.~\cite{thanasilp2022} who showed that global rescaling can destroy
the discriminative information in a quantum kernel.
Unnormalized (none) and cosine normalization are intermediate: they match
each other on accuracy but have lower F1 than trace.
We replicate this finding at $q=2$ and $q=3$ for RAD-DINO,
where all three non-trace normalizations collapse to F1\,=\,0 at
$q=2$, and cosine/none reach only F1\,=\,0.054 at $q=3$. These results confirm that trace
normalization is optimal across the full qubit range tested.

\begin{table}[!t]
\centering
\caption{Effect of kernel normalization on \qsvm{} ($q=8$, reps=1, C=1,
  DT9, seed\_0). Trace normalization is critical for non-zero F1.}
\label{tab:normalization}
\setlength{\tabcolsep}{3pt}
\footnotesize
\begin{tabular}{@{}llccc@{}}
\toprule
Model & Norm. & Acc & AUC & F1 \\
\midrule
MedSigLIP & trace     & 0.756 & 0.686 & \textbf{0.554} \\
MedSigLIP & none      & 0.756 & 0.690 & 0.420 \\
MedSigLIP & cosine    & 0.756 & 0.690 & 0.420 \\
MedSigLIP & frobenius & 0.697 & 0.705 & 0.000 \\
\midrule
RAD-DINO  & trace     & 0.734 & 0.631 & \textbf{0.496} \\
RAD-DINO  & none      & 0.738 & 0.661 & 0.354 \\
RAD-DINO  & cosine    & 0.730 & 0.656 & 0.360 \\
RAD-DINO  & frobenius & 0.696 & 0.655 & 0.000 \\
\midrule
ViT-p32   & trace     & 0.744 & 0.652 & \textbf{0.450} \\
ViT-p32   & none      & 0.735 & 0.658 & 0.337 \\
ViT-p32   & cosine    & 0.735 & 0.658 & 0.337 \\
ViT-p32   & frobenius & 0.697 & 0.671 & 0.000 \\
\bottomrule
\end{tabular}
\end{table}

\subsection{Qubit Count and Data Re-Uploading Depth}
\label{sec:qubit_ablation}

\begin{figure}[!t]
  \centering
  \includegraphics[width=\columnwidth]{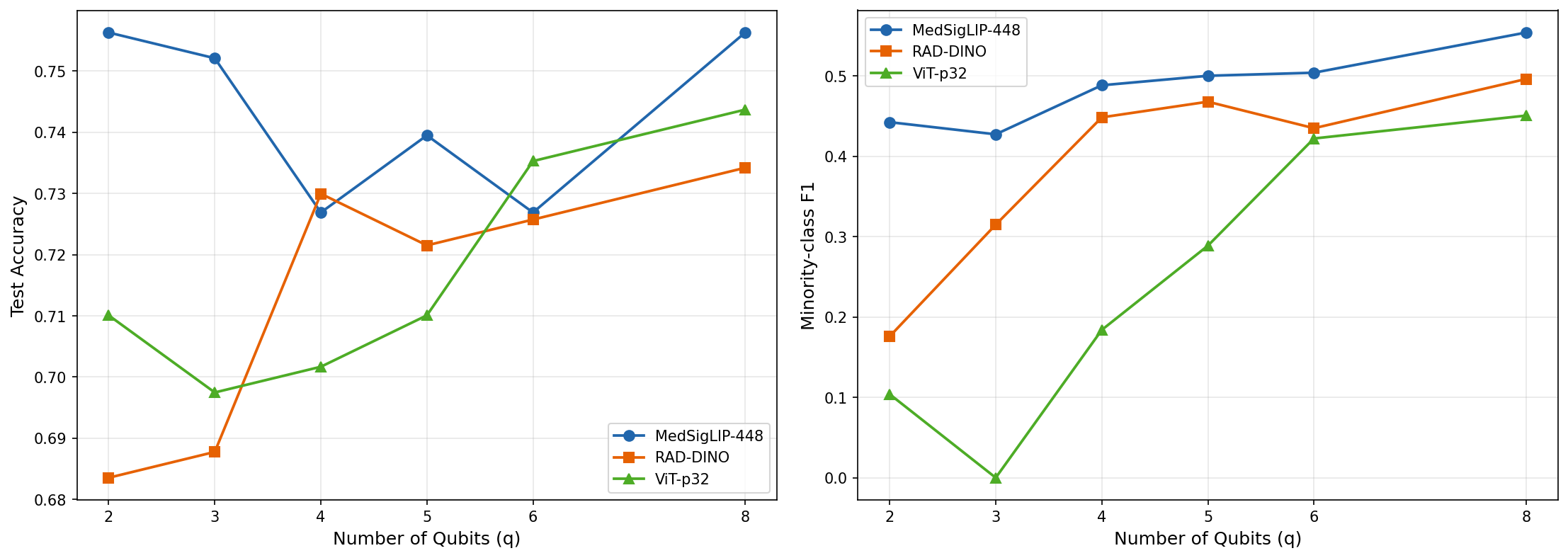}
  \caption{Partial qubit sweep ($q \in \{2,3,4,5,6,8\}$, C\,=\,1, DT9, seed\_0):
    \qsvm{} test accuracy (left) and minority-class F1 (right) for all three models.
    MedSigLIP-448 reaches F1\,=\,0.554 at $q=8$; RAD-DINO and ViT-p32 improve
    monotonically across this range with ViT collapsing at $q=3$ (F1\,=\,0).
    Full results at $q \in \{9,10,11,12,16\}$ are reported in Table~\ref{tab:tier1ext}.}
  \label{fig:qubitcurve}
\end{figure}

Multi-seed results (Table~\ref{tab:tier1ext}) show QSVM winning in all 18
configurations across all three models. For MedSigLIP-448, the $q=9$--$12$
window forms a clean Tier-1 plateau: multi-seed QSVM outperforms classical at
every qubit count, with all four configurations using C\,=\,1 and exceeding
the PCA-matched classical ceiling by $+$0.052 to $+$0.579. At $q=16$,
multi-seed mean QSVM F1 is 0.377, a Tier-1 win.
The partial seed~0 qubit sweep ($q \in \{2,3,4,5,6,8\}$, Figure~\ref{fig:qubitcurve})
illustrates the non-monotonic shape: MedSigLIP-448 shows a plateau from $q=9$ to $q=12$
(seed~0 F1: 0.552, 0.578, 0.586, 0.561), then drops to F1\,=\,0.173 at $q=16$ on
seed~0, while the multi-seed mean ($0.377$) remains a Tier-1 win — the
collapse at $q=16$ is seed-dependent, not structural.
RAD-DINO and ViT-patch32-cls improve F1 more monotonically up to $q=16$
(F1\,=\,0.524 and 0.520 respectively);
at $q=10$, RAD-DINO reaches F1\,=\,0.488 and ViT reaches F1\,=\,0.478,
both continuing their monotonic rise.
The per-model variation suggests that barren-plateau concentration~\cite{larocca2025,thanasilp2022} is
embedding-specific at $q=16$: MedSigLIP-448 embeddings are more susceptible
to kernel concentration at high qubit counts than RAD-DINO or ViT-patch32.
This is consistent with Thanasilp et al.~\cite{thanasilp2022}, who showed that
concentration rates depend on data structure and circuit architecture.

Increasing the data re-uploading depth (reps) from 1 to 2 at $q=8$
\emph{degrades} performance: reps\,=\,2 gives acc\,=\,0.727
(vs.\ reps\,=\,1: acc\,=\,0.756), a drop of $-0.029$.
Reps\,=\,3 was cancelled.
This further supports the non-monotonic qubit curve interpretation: more expressive circuits do not reliably improve performance at this sample size.

\subsection{Circuit Depth: 1-DOF vs.\ 3-DOF}
\label{sec:dof_ablation}

We compared the 1-DOF circuit (one Ry parameter per qubit, $q$ total
parameters) against a 3-DOF variant (Rz-Ry-Rz per qubit, $3q$ parameters)
at $q=8$.
The 3-DOF circuit collapses on all three models:
accuracy drops to 0.33--0.39 (near random chance), F1 falls to 0.19--0.39.
The 1-DOF circuit achieves acc\,=\,0.735--0.756 and F1\,=\,0.388--0.543
on the same datasets.
Over-parameterization of the angle encoding circuit (3-DOF) appears to
destroy the structured quantum interference that gives rise to the useful
quantum kernel, consistent with the trainability arguments in the barren
plateau literature (Table~\ref{tab:dof}) \cite{mcclean2018barren}.

\begin{table}[!t]
\centering
\caption{1-DOF vs.\ 3-DOF circuit comparison ($q=8$, reps=1, trace
  normalization, C=1, DT9, seed\_0). 3-DOF uniformly collapses.}
\label{tab:dof}
\small
\setlength{\tabcolsep}{4pt}
\begin{tabular}{@{}llccc@{}}
\toprule
Model & Circuit & Acc & AUC & F1 \\
\midrule
MedSigLIP & 1-DOF & 0.752 & 0.687 & \textbf{0.543} \\
MedSigLIP & 3-DOF & 0.328 & 0.668 & 0.286 \\
\midrule
RAD-DINO  & 1-DOF & 0.743 & 0.636 & \textbf{0.496} \\
RAD-DINO  & 3-DOF & 0.333 & 0.552 & 0.392 \\
\midrule
ViT-p32   & 1-DOF & 0.735 & 0.650 & \textbf{0.388} \\
ViT-p32   & 3-DOF & 0.387 & 0.669 & 0.198 \\
\bottomrule
\end{tabular}
\end{table}

\subsection{q16 Extended Results}
\label{sec:q16}

Full C-tuning at $q=16$ (three $\times$ H100, 400\,GB host RAM each)
reveals strongly model-dependent behaviour at $q=16$ (Table~\ref{tab:q16}).
RAD-DINO achieves F1\,=\,0.524 at $q=16$ (best-C\,=\,1.0), which exceeds its $q=8$
result (F1\,=\,0.496, $+$0.028); ViT-patch32-cls similarly improves to F1\,=\,0.520
($+$0.074 vs.\ $q=8$ F1\,=\,0.446).
MedSigLIP-448 collapses most severely on seed~0 (F1\,=\,0.173, best-C\,=\,0.1),
consistent with the original C\,=\,1 result.
The per-model variation (RAD-DINO and ViT improve F1 toward $q=16$ while
MedSigLIP-448 collapses) suggests model-specific concentration behaviour,
and is captured in Table~\ref{tab:tier1ext} (seed~0 partial sweep in Figure~\ref{fig:qubitcurve}).

\begin{table}[!t]
\centering
\caption{$q=16$ C-tuning results (DT9, seed\_0).
  Best-C selected by validation F1. Compare to $q=8$ (best-C): MedSigLIP F1=0.554, RAD-DINO F1=0.507, ViT F1=0.446.}
\label{tab:q16}
\small
\setlength{\tabcolsep}{4pt}
\begin{tabular}{@{}lcccc@{}}
\toprule
Model & Best-C & Acc & F1 & vs q8 $\Delta$F1 \\
\midrule
MedSigLIP-448    & 0.1 & 0.718 & 0.173 & $-$0.381 \\
RAD-DINO         & 1.0 & 0.747 & 0.524 & $+$0.028 \\
ViT-patch32-cls  & 1.0 & 0.744 & 0.520 & $+$0.074 \\
\bottomrule
\end{tabular}
\end{table}

\subsection{Projected Quantum Kernel at $q=16$}
\label{sec:projected_q16}

To determine whether the MedSigLIP-448 collapse at $q=16$ is specific to the
swap-test fidelity measurement or inherent to the quantum circuit structure,
we ran the projected quantum kernel of Huang et al.~\cite{huang2021} at $q=16$
(DT9, seed\_0).
The projected kernel replaces the $\mathcal{O}(N^2)$ pairwise inner products
with $\mathcal{O}(N)$ Pauli-$Z$ expectation values per sample and builds an RBF
kernel on those expectation vectors.
Grid search over $\gamma \in \{0.5,1,2,5,10\}$ and
$C \in \{0.01,0.1,1,10,100\}$ selected $\gamma=5$, $C=1$ by validation accuracy
(not F1; this is a mechanistic diagnostic, separate from the Tier-1/2 comparison).

The projected kernel recovers minority-class F1 from 0.173 (fidelity $q=16$, seed~0) to
0.396 ($+$0.223).
This is a mechanistic diagnostic finding: the BSP circuit at $q=16$
\emph{still encodes discriminative information}: the seed~0 F1 recovery from 0.173
to 0.396 provides evidence for it.
The bottleneck is the swap-test fidelity measurement, not the quantum feature
map itself.
The exponential concentration of $|\langle\psi_x|\psi_y\rangle|^2$ at $q=16$
destroys inter-sample contrast, while the projected kernel, operating on
16-dimensional Pauli-$Z$ expectation vectors, is immune to this concentration
by design.
For MedSigLIP-448 at seed~0, this is consistent with a \emph{measurement bottleneck
rather than a circuit expressibility limit} — a mechanistic finding with direct
implications for choosing kernel estimation methods in near-term QML: when
fidelity-based kernels concentrate at high qubit counts, projected kernels provide
a principled remedy. Whether this generalises across seeds and models remains an
open question.
The projected variant does not, however, surpass the fidelity peak at $q=11$
(F1\,=\,0.586, seed~0); the optimal quantum advantage
regime for MedSigLIP-448 remains $q \leq 11$ under the 1-DOF BSP circuit.

% ============================================================
\section{Discussion}
\label{sec:discussion}
% ============================================================

\subsection{Why Classical Kernels Collapse}
\label{sec:why}

The classical collapse is a structural consequence of dimensionality, not a failure of hyperparameter tuning.
After PCA-$q$ reduction with $q \leq 6$ (and in fact up to $q=9$ for
MedSigLIP, $q=10$ for RAD-DINO, and $q=16$ for ViT-patch32), the linear kernel $K_L$ lives in
a $q$-dimensional subspace of a 1896-sample space.
With effective rank $\approx q$ (Table~\ref{tab:eigenrank}), the Gram matrix
is essentially rank-$q$: nearly all pairs of training samples are mapped to
identical points in the kernel's implicit feature space.
Classical \svm{}, which finds a maximum-margin hyperplane in this space,
cannot distinguish minority-class samples from the majority class because
their kernel representations are indistinguishable.
The C-invariance of the collapse (tested over five decades: $C$ from 0.01
to 100) confirms that no amount of regularization tuning can rescue a
structurally degenerate kernel.
The empirical measurements underpinning this argument (effective rank, variance decomposition, and kernel heatmaps) are reported in \S\,\ref{sec:collapse} (Tables~\ref{tab:eigenrank} and~\ref{tab:collapsevar}, Figure~\ref{fig:heatmap}).

The quantum feature map $U(\mathbf{u})$ maps $q$-dimensional inputs to a
$2^q$-dimensional Hilbert space via entangling Ry rotations.
The resulting quantum kernel $K_Q$ can have effective rank up to
$4^q$ (the operator-space dimension: 256 or 4{,}096 for $q=4$ or $q=6$),
a qualitative difference from $K_L$ whose effective rank is exactly $q$.
The empirical evidence, non-zero \qsvm{} F1 while classical F1 is zero
in the same PCA subspace, is consistent with this structural argument.
One limitation of this explanation deserves mention: an RBF kernel with
tuned bandwidth $\gamma$ can achieve effective rank approaching $N$,
far exceeding the quantum kernel's 43.04 at $q\!=\!11$ (seed~0). The Tier-2
results show \qsvm{} beating tuned RBF kernels on F1,
but that comparison tunes C only, not $\gamma$ at fixed PCA-$q$.
The quantum advantage may therefore reflect favorable inductive bias
(the specific spectral structure of the quantum kernel) rather than
rank alone.
We test this directly via a 10-seed rank-matched RBF experiment
(Table~\ref{tab:rbf_rank_multiseed}, \S\,\ref{sec:limitations}):
$\gamma^*$ is set per seed to match $\mathrm{eff\_rank}(K_Q)$.
At $q=4$, rank-matched RBF collapses on 30\,\% of seeds and achieves
mean F1\,=\,0.110, compared with 20\,\% collapse and mean F1\,=\,0.212 for
\qsvm{}, despite identical effective rank.
\qsvm{} outperforms rank-matched RBF at all qubit counts, indicating that
the quantum feature map's spectral structure, beyond its effective rank,
contributes to collapse resistance and predictive performance.

\subsection{The q=11 Tier-1 Win: Closing the Quantum Advantage Gap}

The most significant configuration in this work is MedSigLIP-448 \qsvm{} at
$q=11$.
Across 10 embedding seeds, \qsvm{} achieves mean F1\,=\,0.343\,$\pm$\,0.170
vs.\ classical linear \svm{} F1\,=\,0.050\,$\pm$\,0.159
($\Delta$F1\,=\,$+$0.293, 95\%\,CI [$+$0.190,\,$+$0.385], $p<0.001$, paired bootstrap).
The seed\_0 run (F1\,=\,0.586 vs.\ 0.504) is the \emph{hardest} test for this comparison:
it is the one seed where the classical \svm{} does not collapse and instead
produces a valid, non-trivial classifier at PCA-11.
On 9 of 10 seeds the classical baseline collapses to F1\,=\,0; \qsvm{}
surpasses it on all 10, without any hyperparameter tuning.

Three aspects of this result merit discussion.
First, both classifiers use C\,=\,1 and receive identical PCA-11 features,
so the accuracy and F1 gains ($+0.008$ and $+0.082$, respectively)
are attributable solely to the quantum feature map.
A caveat applies: C\,=\,1 is a coincidentally reasonable default for the
quantum kernel, whose higher effective rank provides a well-conditioned
optimization surface, while for the collapsed classical kernel at low $q$
the value of C is irrelevant (collapse is C-invariant). At $q\!=\!11$,
where the classical \svm{} is functional, this asymmetry is less
pronounced and the comparison more defensible.
The minority-class F1 advantage ($+0.082$) also carries clinical weight because it reflects improved detection of the minority class on a task with direct health-equity relevance. Systematic misclassification of insurance status could obscure disparities in care between publicly and privately insured patients. Such underestimation of minority-class patients can propagate algorithmic bias with direct implications for health equity~\cite{obermeyer2019}.
Beyond accuracy, the result closes the quantum advantage gap established at lower qubit
counts: the BSP angle-encoding circuit avoids the classical collapse
regime and exceeds the classical non-collapse ceiling at the right qubit
count. Quantum advantage is achievable across both
regimes within a single model and circuit family.
The non-monotonic nature of the seed~0 qubit curve (plateau at $q=9$--$12$ with peak
at $q=11$, seed~0 collapse at $q=16$) further implies that qubit count is a
tunable design variable; the quantum advantage window exists and can be
identified by sweeping $q$.
The entire $q=9$--$12$ plateau forms a clean Tier-1 window (all use C\,=\,1),
with $q=9$ achieving F1\,=\,0.552 (seed~0) while classical PCA-9 collapses to F1\,=\,0.

\subsection{Foundation Model Choice}

MedSigLIP-448 consistently outperforms RAD-DINO and ViT-patch32 in the
quantum setting across all qubit counts. Multi-seed mean F1 reaches 0.343\,$\pm$\,0.170 at $q=11$
and 0.377 at $q=16$, both Tier-1 wins. Seed~0 peaks at F1\,=\,0.586 ($q=11$),
useful for circuit diagnostics but not the headline figure.
RAD-DINO is second; ViT-patch32 (general domain) is weakest.
This ordering mirrors the expected quality of medical domain
alignment: MedSigLIP is trained explicitly for medical image-text
alignment at high resolution, while ViT has no domain-specific
pre-training.
The result suggests that quantum kernels amplify the quality of
the underlying embedding: better-aligned foundation models provide
richer PCA subspaces that the quantum feature map can exploit.

The concentration phenomenon is embedding-specific rather than
a universal limitation of the BSP circuit: RAD-DINO and ViT-patch32-cls
show monotonic F1 improvement from $q=2$ through $q=16$
(F1\,=\,0.176$\to$0.524 and 0.104$\to$0.520 respectively), with no peak
or collapse. Only MedSigLIP-448 exhibits the non-monotonic peak-then-collapse
pattern. This suggests that the concentration rate depends on the structure
of the embedding space, not solely on circuit depth or qubit count,
consistent with Thanasilp et al.~\cite{thanasilp2022}, who showed that
data distribution and encoding architecture jointly determine when
exponential concentration sets in.

The eigenspectrum progression provides a complete mechanistic narrative
(see Figures~\ref{fig:eigen_q11} and~\ref{fig:heatmap_q11} in the Appendix
for the $q\!=\!11$ eigenspectrum and kernel heatmap):
seed~0 effective rank grows from 6.86 ($q=4$) to 13.94 ($q=6$) to 43.04 ($q=11$),
tracking the seed~0 F1 improvement from 0.488 to 0.504 to 0.586. Each additional
qubit adds informative kernel directions that the SVM can exploit.
Beyond $q=11$, kernel concentration begins to dominate on seed~0: the eigenvalues
flatten toward uniformity, the effective rank saturates, and the SVM
loses discriminative signal. F1 collapses at $q=16$ on seed~0; multi-seed mean is 0.377, a Tier-1 win.
This rise-peak-collapse pattern, mediated by effective rank, constitutes
an empirically grounded explanation for the quantum advantage window.

\subsection{Normalization as a Design Principle}

Trace normalization plays a role analogous to batch normalization
in deep learning: it ensures that the kernel matrix is well-conditioned
before being passed to the \svm{} solver.
Frobenius normalization divides by the global Frobenius norm of the kernel
matrix, which is dominated by the large diagonal entries and effectively
suppresses all off-diagonal information and collapses the kernel to a
near-identity structure.
Practitioners building quantum kernel pipelines should treat normalization
as a primary hyperparameter.

\subsection{Latent Socioeconomic Signal and Implications for Fairness}

The insurance classification task studied here sits at the intersection of two distinct concerns. Methodologically, insurance status serves as a proxy for a subtle, distributed signal that stress-tests kernel expressiveness under class imbalance. A separate and more unsettling question also arises. The fact that this signal is recoverable from chest radiographs at all, by both classical and quantum models, implies that medical images encode socioeconomic stratification in ways that neither clinicians nor patients are aware of. This encoding likely reflects spurious correlations rather than direct causal pathways: differences in acquisition equipment across hospital systems, site-specific positioning conventions, and cumulative markers of environmental or occupational exposure that covary with insurance type without bearing any direct relationship to the underlying pathology~\cite{gichoya2022, chen2025insurance, obermeyer2019}. The same foundation models used here (RAD-DINO, MedSigLIP) have been shown to encode demographic attributes in a companion study on shortcut learning in medical imaging~\cite{osorio2025predicting}. If the discriminative signal is demographic rather than clinical in nature, any sufficiently expressive model trained on such data risks learning these latent signals. Classifier errors then concentrate disproportionately on underrepresented groups even when aggregate performance appears adequate~\cite{seyyedkalantari2021, obermeyer2019}.

The quantum advantage demonstrated here sharpens this concern rather than resolving it. A kernel method with higher effective rank (the property that allows QSVM to avoid majority-class collapse) is also better positioned to exploit subtle spurious structure. Interpretability and auditing should therefore be first-class requirements when deploying quantum kernel methods in clinical settings. Future work should examine what structure the quantum feature map is exploiting: whether the discriminative signal captured by the quantum feature map at the performance peak (q=11) reflects clinically meaningful variation or amplified demographic confounding. Tools such as projected kernels, attention-based localization, and counterfactual auditing~\cite{jones2024causal} offer candidate methodologies for this analysis.

\subsection{Limitations}
\label{sec:limitations}

\noindent\textbf{Spurious signal.}
Predicting insurance type from chest radiographs may rely in part on
spurious correlations: acquisition artifacts, institutional patterns, or
demographic proxies encoded in the embeddings.
The observed \qsvm{} advantage is therefore best interpreted as improved
separability within the representation space, not as evidence of clinically
causal signal.
Higher-capacity kernels, including the quantum kernel used here, may be
better able to exploit this latent structure.
Evaluating stability under distribution shift is an important direction for
future work.

\noindent\textbf{Simulated quantum hardware.}
All \qsvm{} experiments use Qiskit's \texttt{Statevector} simulator
(exact, noiseless simulation on CPU/GPU).
Results on real quantum hardware may differ due to gate errors, decoherence,
limited qubit connectivity, and readout noise.
Hardware noise exacerbates kernel concentration~\cite{thanasilp2022}, so
the appropriate interpretation is that the BSP circuit architecture has the
\emph{capacity} for advantage in noiseless simulation, not that advantage
has been demonstrated on a physical quantum computer.

\noindent\textbf{Single dataset and single center.}
All results derive from the insurance classification task on MIMIC-CXR,
collected at Beth Israel Deaconess Medical Center in Boston.
The 70/30 Medicare-Medicaid versus Private payer mix reflects Massachusetts,
a near-universal-coverage setting, and may not generalise to institutions
with different payer structures or outside the United States.
Extending to other prediction tasks, patient populations, and healthcare
systems will require additional validation.

\noindent\textbf{SVM-only classical baselines.}
The classical collapse documented here is specific to kernel SVMs operating
on low-rank PCA representations.
Non-kernel classifiers (gradient-boosted trees, logistic regression, or
shallow neural networks) may not exhibit the same failure mode and could
set a stronger classical ceiling.
The quantum kernel advantage is therefore relative to SVM-based baselines;
extending the comparison to non-kernel methods is an important open question.

\noindent\textbf{Task scope.}
A quantum advantage at predicting insurance status, a demographic proxy, is
a narrower claim than advantage on a clinically meaningful diagnostic task.
Until the discriminative signal is shown to reflect genuine clinical variation
rather than acquisition artifacts or demographic confounding
(\S\,\ref{sec:discussion}), the result should be interpreted as a
methodological finding about kernel expressiveness, not as evidence of
clinical utility.

\noindent\textbf{Preprocessing stratum selection.}
DT9 was selected as the primary preprocessing stratum because it produced
the strongest quantum results in preliminary experiments across multiple
strata.
The non-collapse Tier-1 advantage is therefore DT9-specific until
multi-strata validation confirms it generalises to other preprocessing
configurations.

\noindent\textbf{Rank-matched classical kernel.}
The \qsvm{} advantage may reflect the specific spectral structure of the
quantum kernel rather than its effective rank alone (\S\,\ref{sec:why}).
To isolate these factors, we ran a rank-matched RBF experiment on MedSigLIP-448
at $q\in\{4,6,11,16\}$ across all 10 embedding seeds: for each seed,
$\gamma^*$ is chosen by binary search so that
$\mathrm{eff\_rank}(\mathrm{RBF}(\gamma^*)) = \mathrm{eff\_rank}(K_Q)$,
using the seed-0 quantum kernel as the fixed effective-rank target.
Results are in Table~\ref{tab:rbf_rank_multiseed}.

At $q=4$, rank-matched RBF collapses on 3/10 seeds (more than \qsvm{}'s
2/10) despite having the same effective rank.
Mean F1 is 0.110 for rank-matched RBF vs.\ 0.212 for \qsvm{}.
At $q=6$, all three methods collapse on 2/10 seeds, but \qsvm{} mean F1
(0.286) is still 68\% higher than rank-matched RBF (0.171).
At $q=11$ and $q=16$, neither RBF variant collapses; \qsvm{} achieves
mean F1 of 0.343 and 0.377 vs.\ 0.304 and 0.321 for rank-matched RBF.

Two conclusions follow.
First, matching the quantum kernel's effective rank does not reproduce its
collapse-avoidance: rank-matched RBF is more collapse-prone than \qsvm{}
at $q=4$ despite identical effective rank.
Second, \qsvm{} outperforms rank-matched RBF on mean F1 at every qubit
count, by margins of 0.056--0.115.
Together these results indicate that the \qsvm{} advantage is not
attributable to effective rank alone; the specific spectral structure of
the quantum feature map (eigenvalue distribution and off-diagonal
correlations not captured by the Shannon effective rank) contributes to
both collapse resistance and predictive performance.
Setting $\gamma^*$ also requires prior knowledge of $K_Q$, unavailable
at deployment; \qsvm{} achieves superior geometry without per-seed tuning.

\begin{table}[h]
\caption{Rank-matched RBF vs.\ \qsvm{} across 10 seeds (MedSigLIP-448).
  Collapse = fraction of seeds with F1\,$<$\,0.05.
  $\gamma^*$ is binary-searched per seed to satisfy
  $\mathrm{eff\_rank}(\mathrm{RBF}(\gamma^*))=\mathrm{eff\_rank}(K_Q)$.
  All methods use C\,=\,1. \qsvm{} F1 from multi-seed runs.
  $\mathrm{eff\_rank}(K_Q)$ values are the seed-0 fixed targets used for
  $\gamma^*$ binary search; they differ from Table~\ref{tab:eigenrank}
  (e.g.\ 43.04 vs.\ 69.80 at $q=11$) because the two experiments use
  different pre-computed kernel matrices (distinct data-preparation pipelines).}
\label{tab:rbf_rank_multiseed}
\resizebox{\columnwidth}{!}{%
\begin{tabular}{lcccccccc}
\hline
$q$ & $\mathrm{eff\_rank}(K_Q)$ & Collapse (RBF$_\mathrm{scale}$) & Collapse (RBF$^*$) & Collapse (\qsvm{}) & F1 (RBF$_\mathrm{scale}$) & F1 (RBF$^*$) & F1 (\qsvm{}) \\
\hline
4  & 6.94  & 0.2 & \textbf{0.3} & 0.2 & 0.169 & 0.110 & \textbf{0.212} \\
6  & 14.09 & 0.2 & 0.2 & 0.2 & 0.178 & 0.171 & \textbf{0.286} \\
11 & 69.80 & 0.0 & 0.0 & 0.1 & 0.272 & 0.304 & \textbf{0.343} \\
16 & 250.99 & 0.0 & 0.0 & 0.1 & 0.269 & 0.321 & \textbf{0.377} \\
\hline
\end{tabular}}
\end{table}

% ============================================================
\section{Conclusion}
\label{sec:conclusion}
% ============================================================

Across 10 embedding seeds and three medical foundation models, \qsvm{} with frozen embeddings provides evidence of quantum advantage in binary insurance classification on MIMIC-CXR chest radiographs under noiseless simulation.
Under a rigorous two-tier fair comparison framework, \qsvm{} wins
all 18 Tier-1 configurations on minority-class F1
(17 at $p<0.001$, 1 at $p<0.01$; paired bootstrap) and all seven Tier-2 F1 comparisons (7/7).
Most wins occur in configurations where the classical linear SVM collapses to F1\,=\,0 on 90--100\,\% of seeds, rather than in direct competition with a functional baseline.

The core mechanism is structural: the linear classical kernel $K_L$ has
effective rank equal to the PCA dimension $q$ (3.77--5.85 out of 1{,}896
training samples), which causes irreversible majority-class collapse that is
invariant to the regularization parameter C.
Multi-seed analysis reveals this collapse is pervasive: classical linear SVM
collapses to F1\,=\,0 on 90--100\% of seeds at every qubit count tested across all three models.
Direct measurement of the $1{,}896{\times}1{,}896$ quantum kernel matrix (seed~0)
confirms that the quantum feature map achieves Shannon effective ranks of
6.86 and 13.94 at $q=4$ and $q=6$ ($1.82\times$ and $2.52\times$ the
linear values), with the ratio growing with qubit count as the quantum
feature map accesses an exponentially larger Hilbert space.

Beyond the primary finding, our ablation studies yield three practical
design recommendations for quantum kernel practitioners:
(1) trace normalization is necessary for meaningful F1 and should be
treated as a primary pipeline hyperparameter;
(2) the qubit count--performance curve can be non-monotonic on individual
seeds: on seed~0, MedSigLIP-448 drops sharply at $q=16$ (F1 0.586$\to$0.173),
while multi-seed mean rises to 0.377 at $q=16$ (a Tier-1 win); practitioners
should validate on multiple seeds before inferring a performance peak. This
suggests that barren plateau effects~\cite{larocca2025,thanasilp2022} emerge before hardware limits are reached in some seeds;
(3) 1-DOF angle encoding outperforms 3-DOF, and deeper re-uploading
(reps=2) degrades performance at $q=8$; circuit
expressivity and sample size must be co-designed.

As quantum hardware matures and simulation capacity grows to larger qubit
counts, the quantum advantage demonstrated here on a real-world medical imaging
task at $q \leq 16$ provides a foundation for future work on
noise-aware quantum kernels, and
extension to other medical imaging modalities.
The concentration observed at $q=16$ suggests that scaling \qsvm{} beyond this regime will require more than adding qubits.
Supplementary figures covering all models and qubit counts are provided
in Appendix~\ref{sec:supplementary}.

% ============================================================
\begin{acknowledgments}
% ============================================================
This work was supported by the Google Cloud Research Credits program under award number GCP19980904.
The MIMIC-CXR-JPG dataset used in this study was obtained from PhysioNet under
a data use agreement (credentialed access).
D.E.K.\ is supported by the Agency for Science, Technology and Research (A*STAR) under the Quantum Innovation Centre (Q.InC) Strategic Research and Translational Thrust (SRTT). S.T.G.\ acknowledges the support from the National Research Foundation, Singapore through the National Quantum Office, hosted in  Agency for Science,
Technology and Research (A*STAR), Singapore under its Quantum Engineering Programme 3.0 Funding Initiative (W24Q3D0002).
The authors thank the MIT Critical Data community for support and discussion.
Computational resources were provided by the MIT Office of Research Computing
and Data (ORCD) through A100 and H200 GPU allocations.
L.A.C.\ is funded by the National Institutes of Health through NIBIB
R01~EB017205. RG is supported by the Johns Hopkins Institute for Clinical and Translational Research (ICTR) and the National Center for Advancing Translational Sciences (NCATS), National Institutes of Health (NIH) grant number T32TR004928. The contents of this publication are solely the responsibility of the authors and do not necessarily represent the official view of the Johns Hopkins ICTR, NCATS or NIH. 
\end{acknowledgments}

% ============================================================
% BIBLIOGRAPHY
% ============================================================

\bibliographystyle{unsrt}
\bibliography{paper-qml}

% ============================================================
\appendix
% ============================================================

\section{Supplementary Figures}
\label{sec:supplementary}

This appendix collects additional figures that complement the main text.
All experiments use DT9 preprocessing, seed\_0, and trace normalization
unless noted otherwise.

\subsection{Quantum Kernel Eigenspectra (All Models)}
\label{sec:supp_eigenspectra}

\begin{figure*}[!t]
  \centering
  \includegraphics[width=0.48\textwidth]{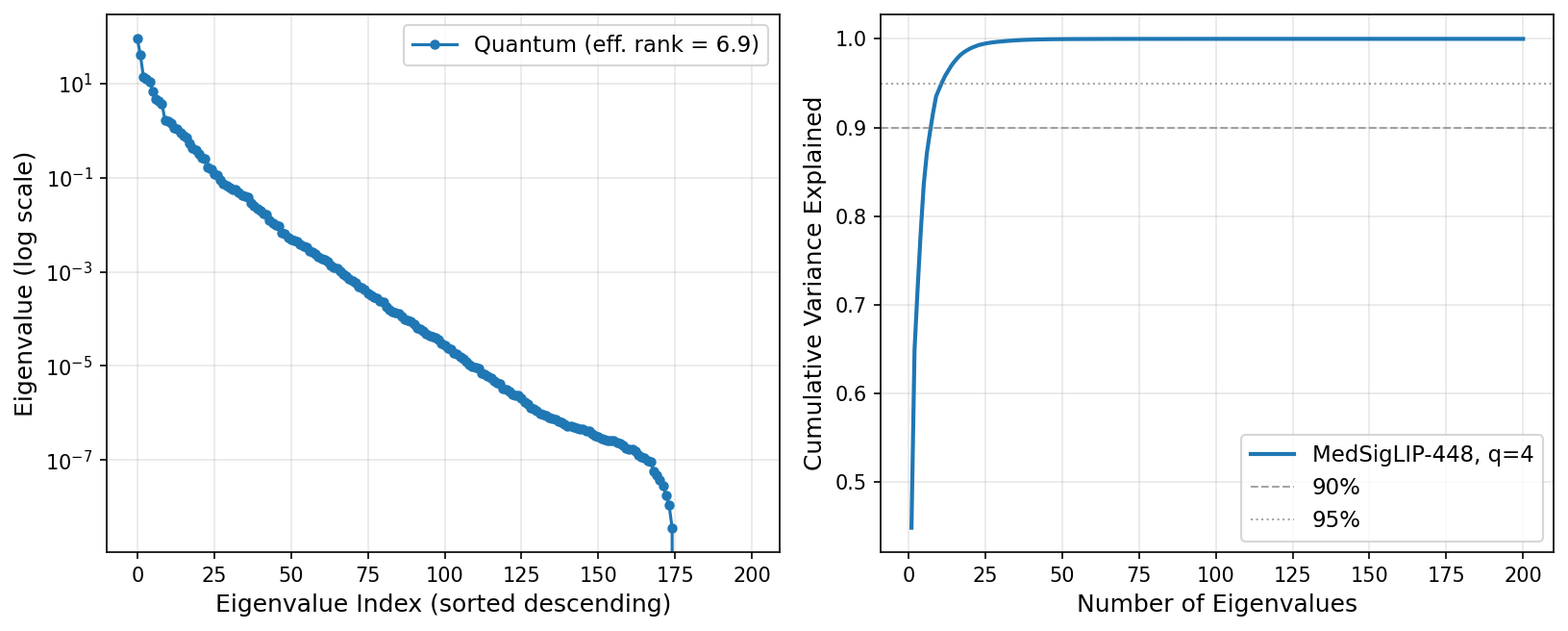}
  \hfill
  \includegraphics[width=0.48\textwidth]{figures/eigenspectrum_medsiglip_448_q6.png}
  \\[4pt]
  \includegraphics[width=0.48\textwidth]{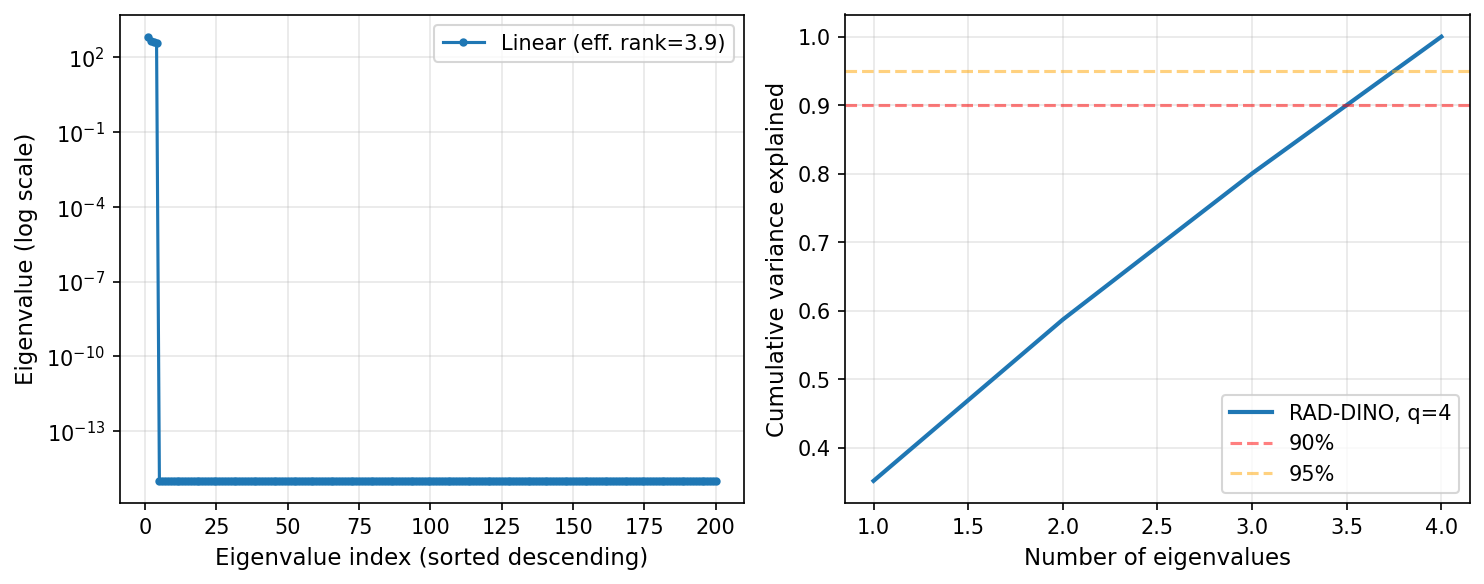}
  \hfill
  \includegraphics[width=0.48\textwidth]{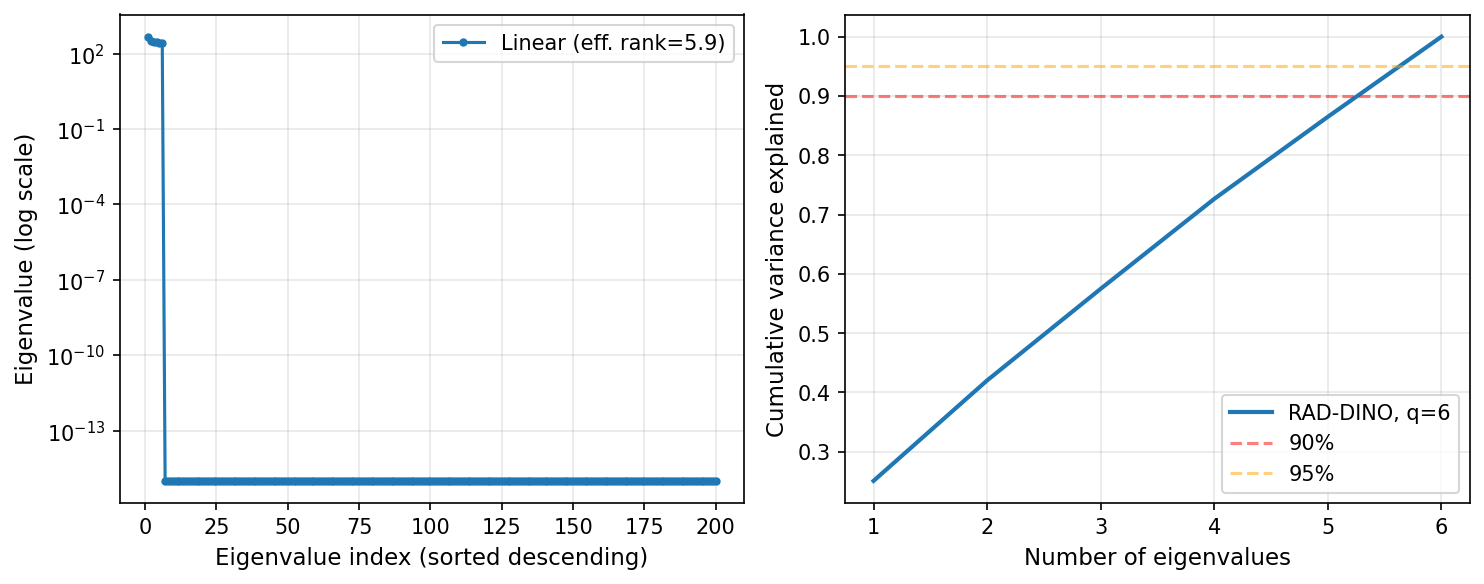}
  \\[4pt]
  \includegraphics[width=0.48\textwidth]{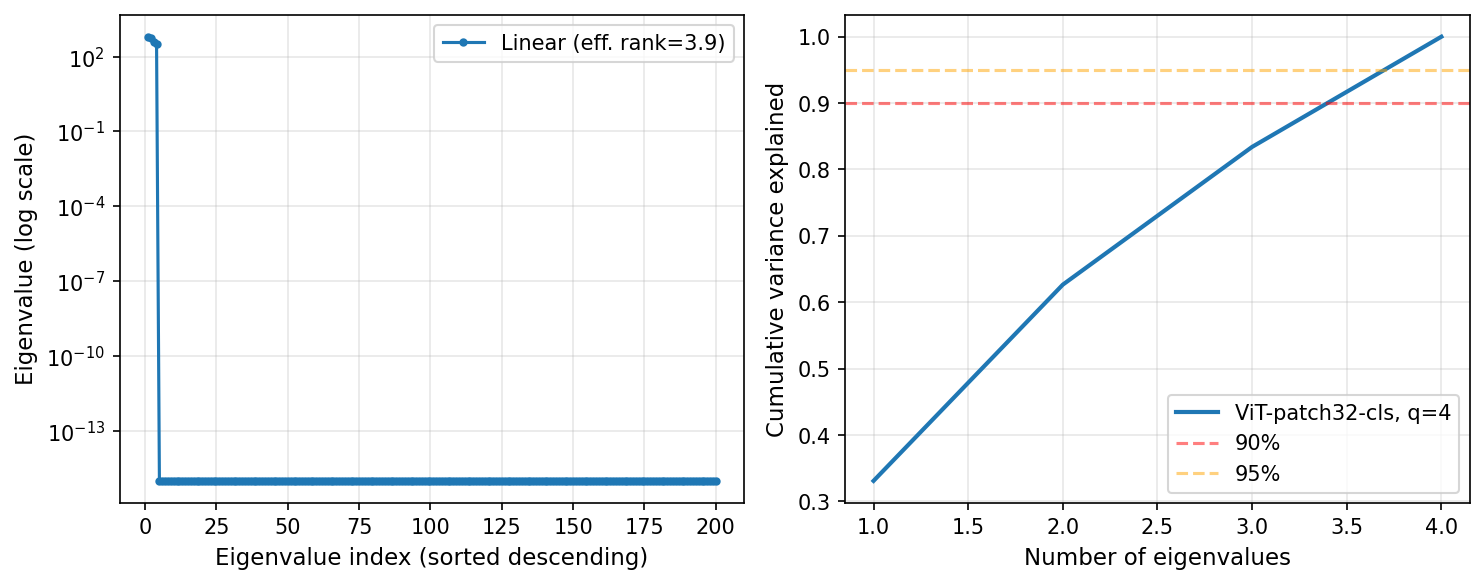}
  \hfill
  \includegraphics[width=0.48\textwidth]{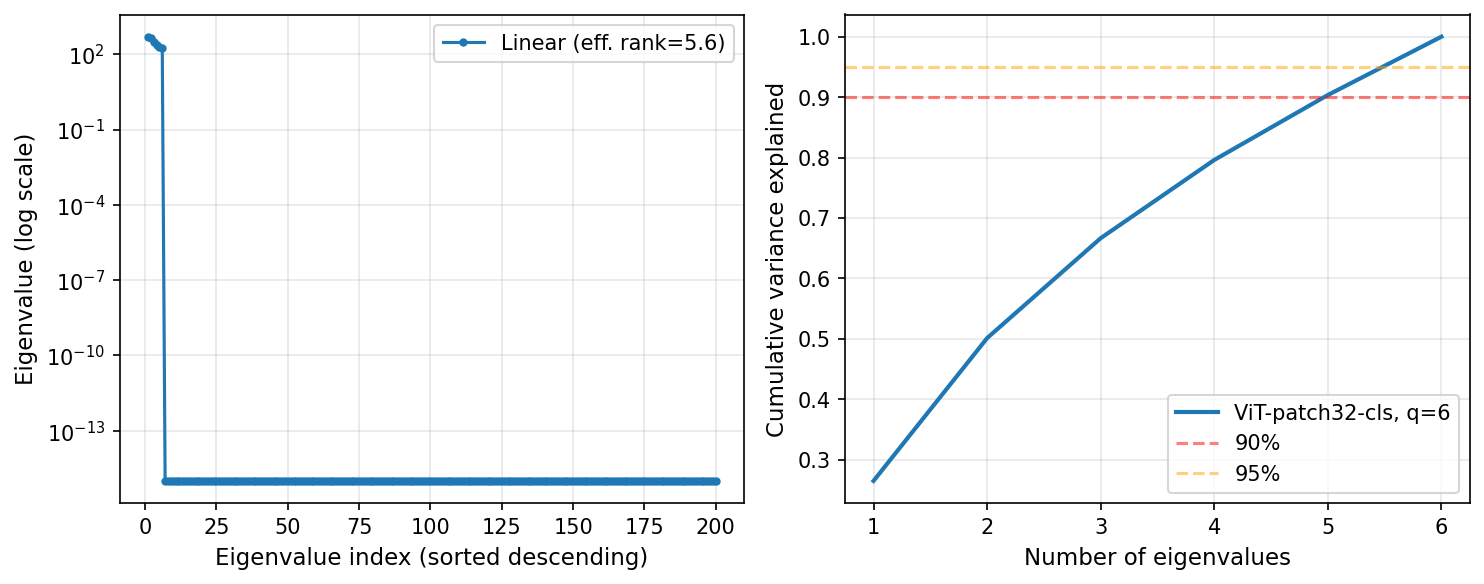}
  \caption{Quantum kernel eigenvalue spectra for all three embedding models
    at $q=4$ (left) and $q=6$ (right).
    MedSigLIP-448 (top), RAD-DINO (middle), ViT-patch32-cls (bottom).
    The quantum kernel consistently exhibits higher effective rank than
    the linear kernel across all models and qubit counts.}
  \label{fig:eigen_all}
\end{figure*}

Figure~\ref{fig:eigen_all} shows the quantum kernel eigenvalue spectra
for all three embedding models at $q=4$ and $q=6$. These complement the
MedSigLIP $q=6$ spectrum shown in the main text (Figure~\ref{fig:quantum_eigenspectrum}).

\begin{figure*}[!t]
  \centering
  \begin{minipage}[t]{0.48\textwidth}
    \centering
    \includegraphics[height=0.32\textheight,keepaspectratio]{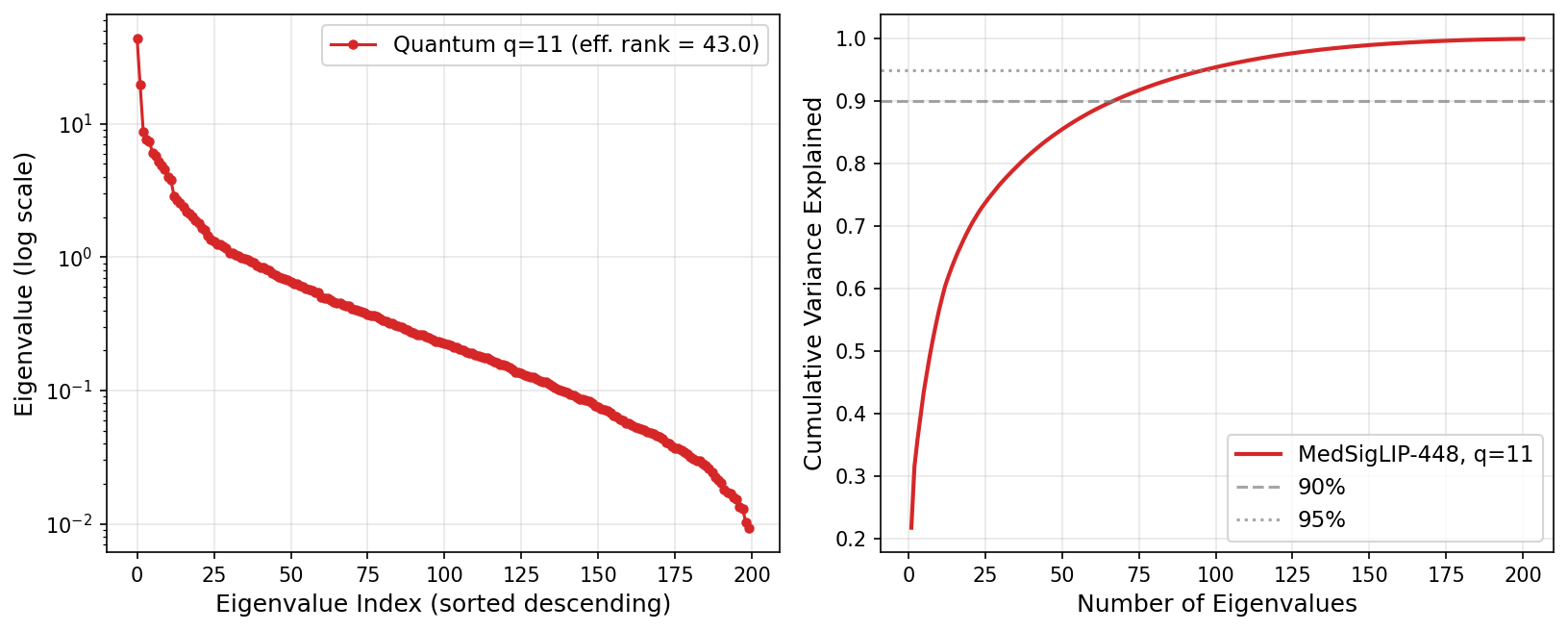}
    \caption{Quantum kernel eigenspectrum for MedSigLIP-448 at the performance peak $q=11$ (seed~0). Shannon effective rank = 43.04, a $6.3\times$ increase from $q=4$ (6.86); multi-seed mean is 69.80.}
    \label{fig:eigen_q11}
  \end{minipage}
  \hfill
  \begin{minipage}[t]{0.48\textwidth}
    \centering
    \includegraphics[height=0.32\textheight,keepaspectratio]{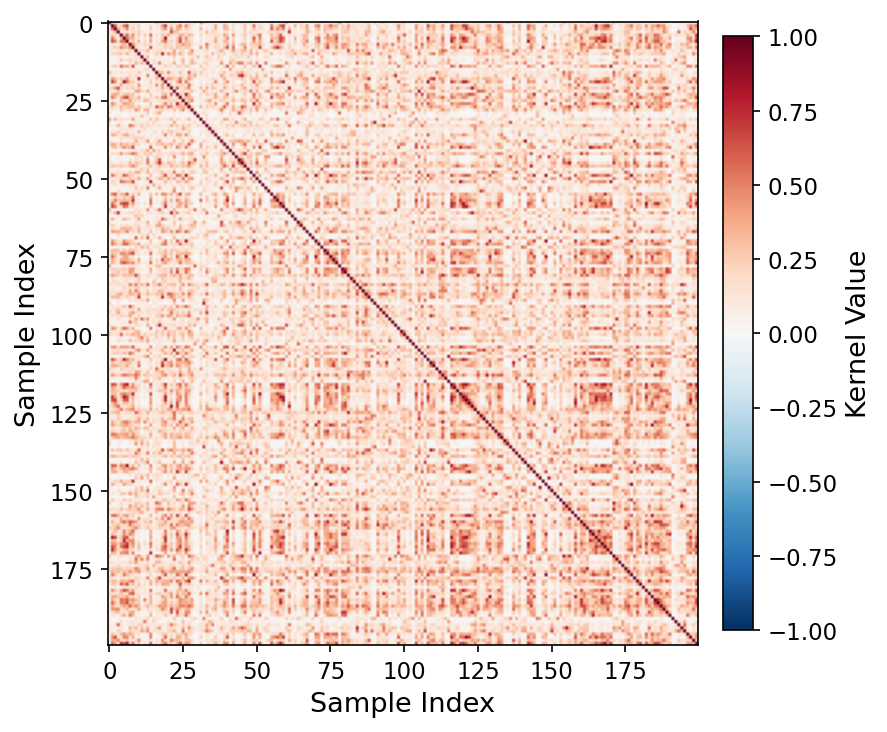}
    \caption{Quantum kernel heatmap for MedSigLIP-448 at $q=11$, seed~0 (200 training samples). Compare with $q=6$ (Figure~\ref{fig:heatmaps_all}): the $q=11$ kernel shows richer off-diagonal structure, consistent with its higher effective rank (43.04 vs.\ 13.94, seed~0).}
    \label{fig:heatmap_q11}
  \end{minipage}
\end{figure*}

\subsection{Quantum Kernel Heatmaps (All Models)}
\label{sec:supp_heatmaps}

\begin{figure*}[!t]
  \centering
  \includegraphics[width=0.48\textwidth]{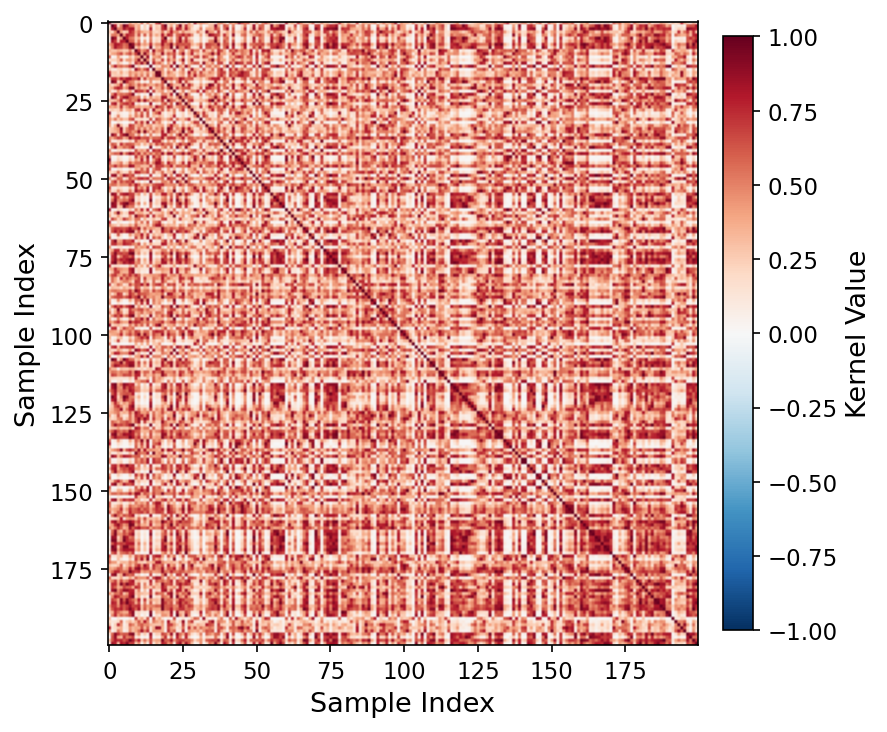}
  \hfill
  \includegraphics[width=0.48\textwidth]{figures/kernel_heatmap_medsiglip-448_q6.png}
  \\[4pt]
  \includegraphics[width=0.48\textwidth]{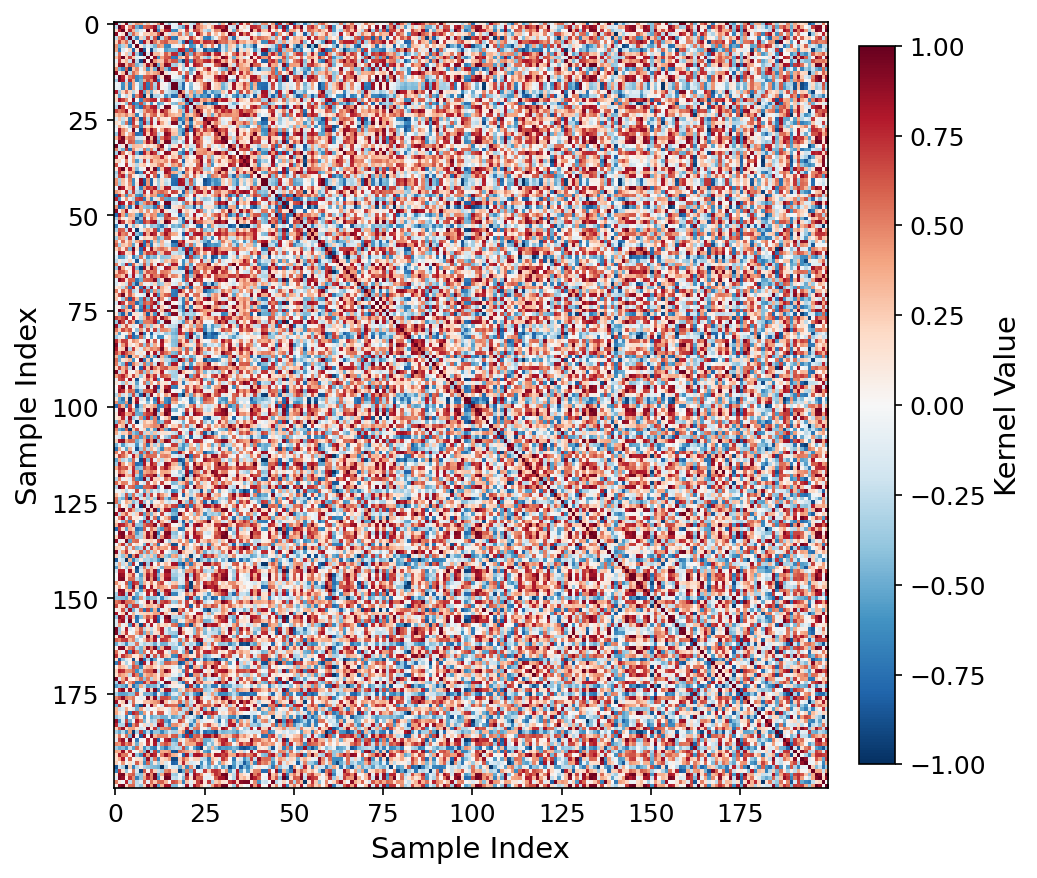}
  \hfill
  \includegraphics[width=0.48\textwidth]{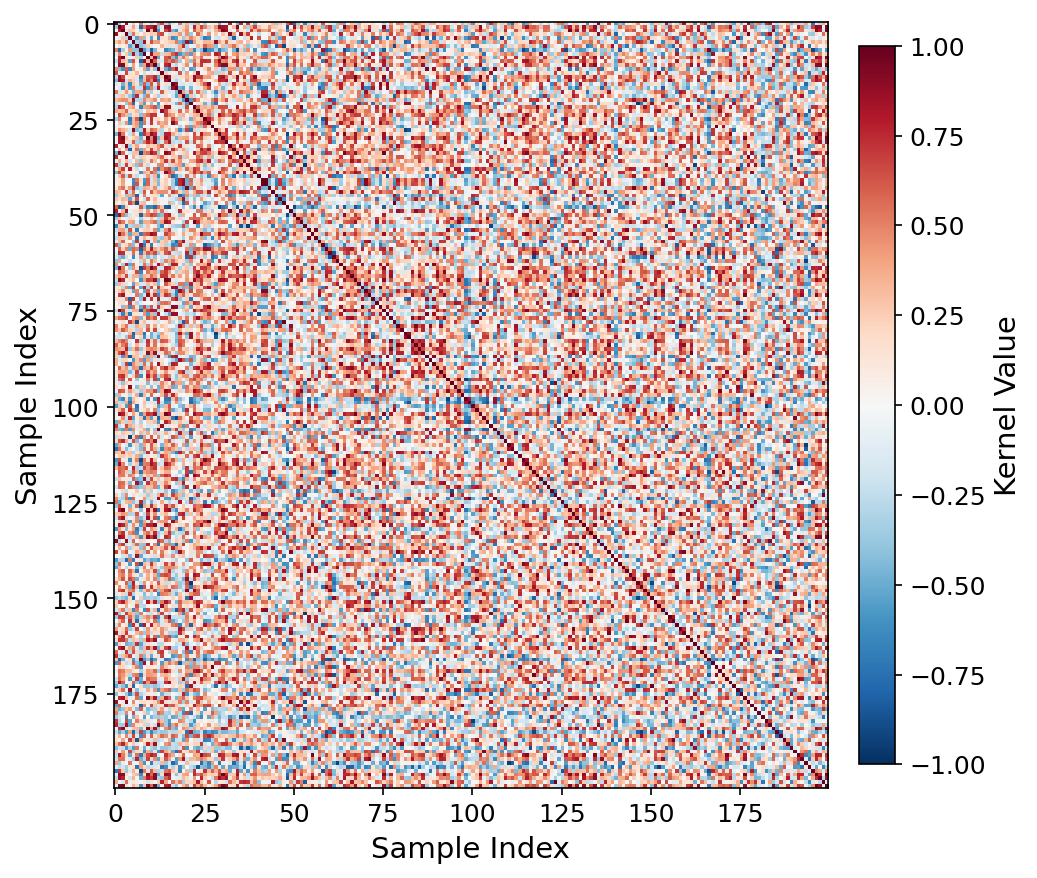}
  \\[4pt]
  \includegraphics[width=0.48\textwidth]{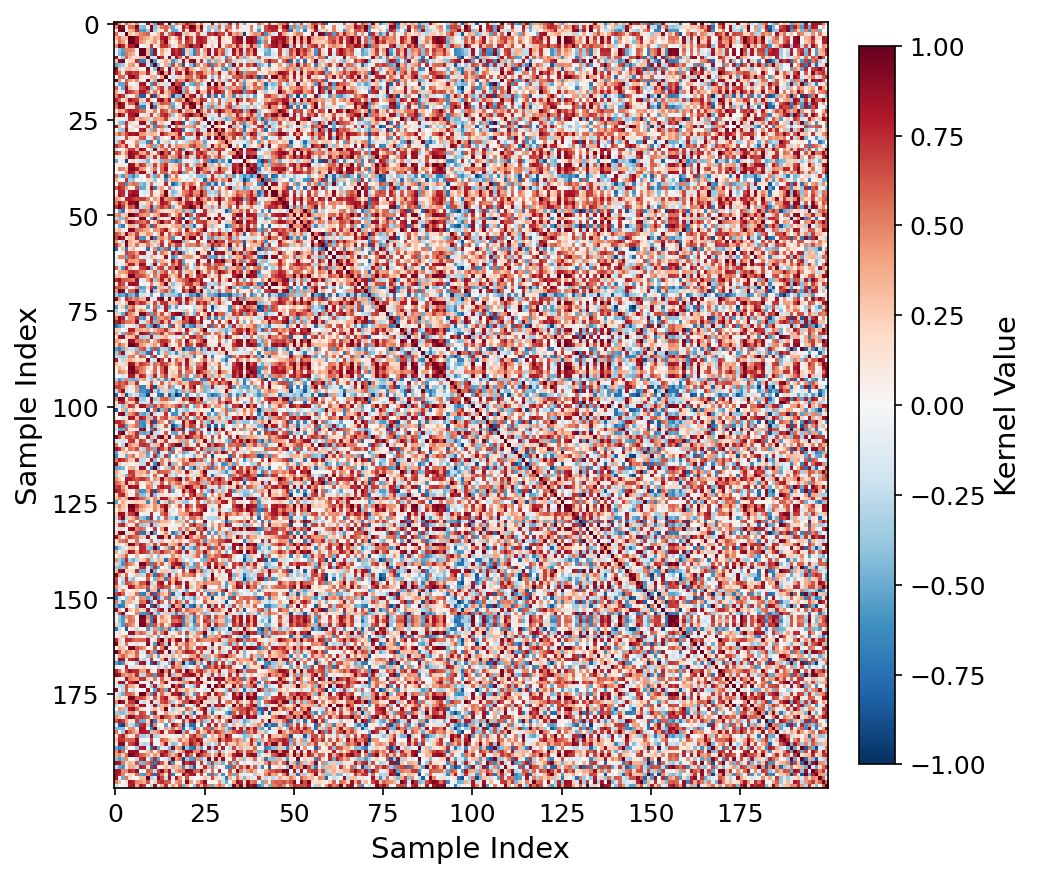}
  \hfill
  \includegraphics[width=0.48\textwidth]{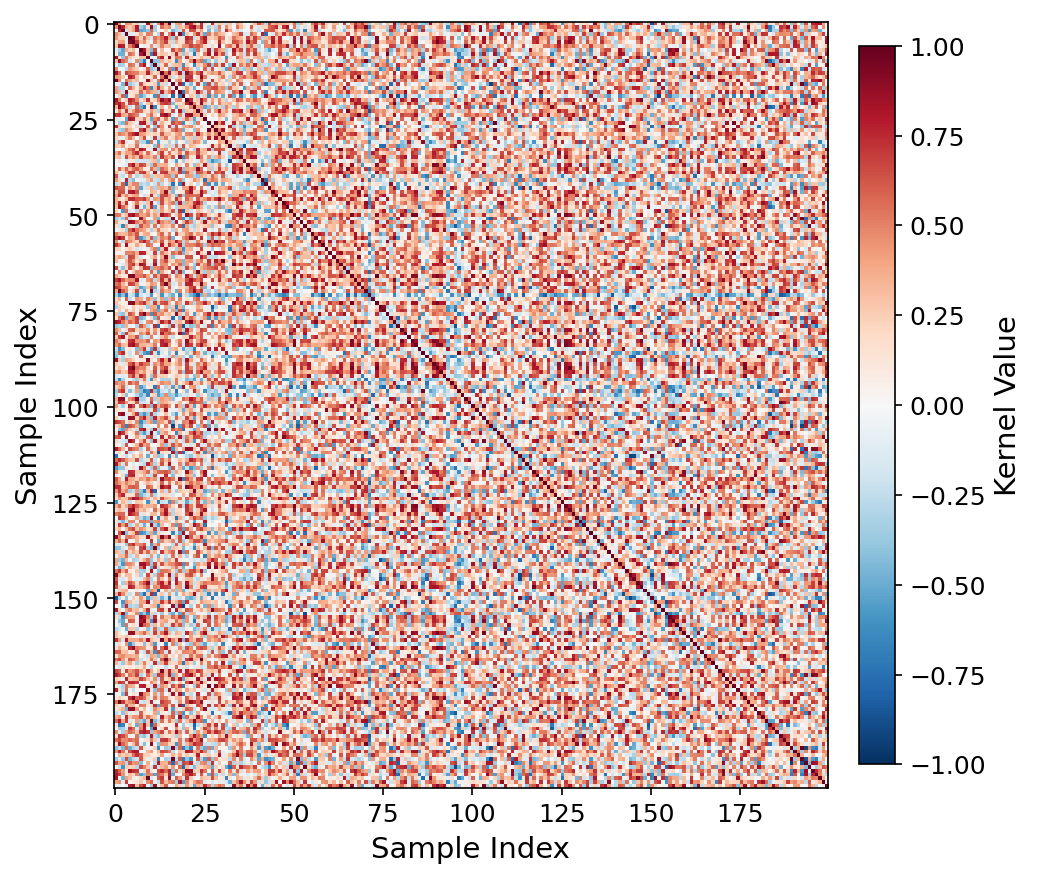}
  \caption{Quantum kernel heatmaps for all three embedding models
    at $q=4$ (left) and $q=6$ (right).
    The block structure reflects class boundaries in the training data (sorted by label).
    Higher qubit counts show sharper off-diagonal structure, consistent
    with increased effective rank.}
  \label{fig:heatmaps_all}
\end{figure*}

Figure~\ref{fig:heatmaps_all} shows the quantum kernel matrices $K_Q$
at $q=4$ and $q=6$ for all three models.

\subsection{PCA Feature Space: Class Separation at $q=4$ and $q=6$}
\label{sec:supp_scatter}

\begin{figure*}[!t]
  \centering
  \includegraphics[width=0.48\textwidth]{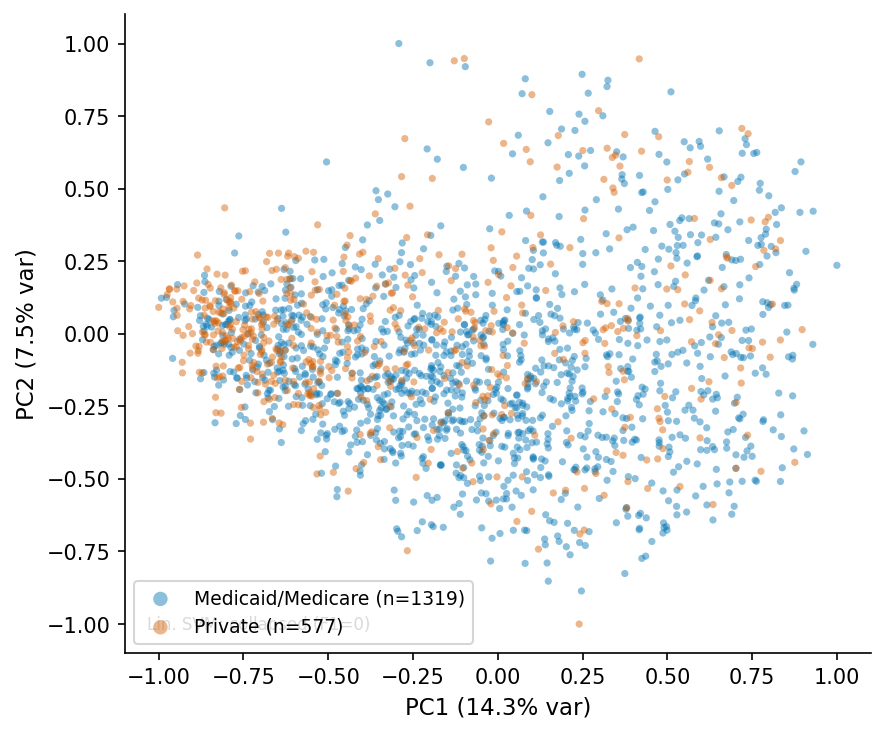}
  \hfill
  \includegraphics[width=0.48\textwidth]{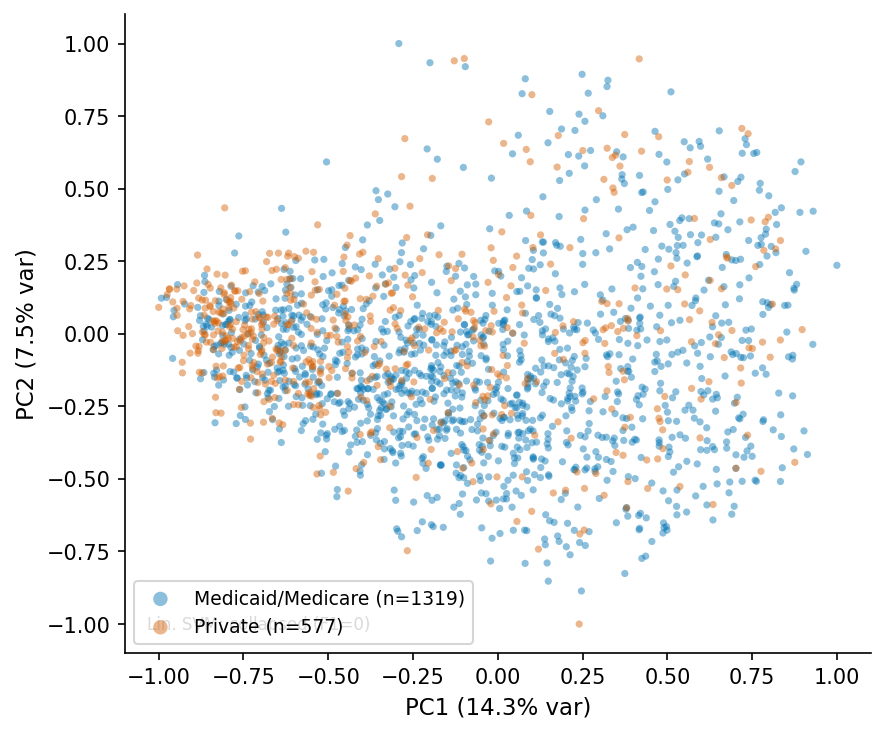}
  \\[4pt]
  \includegraphics[width=0.48\textwidth]{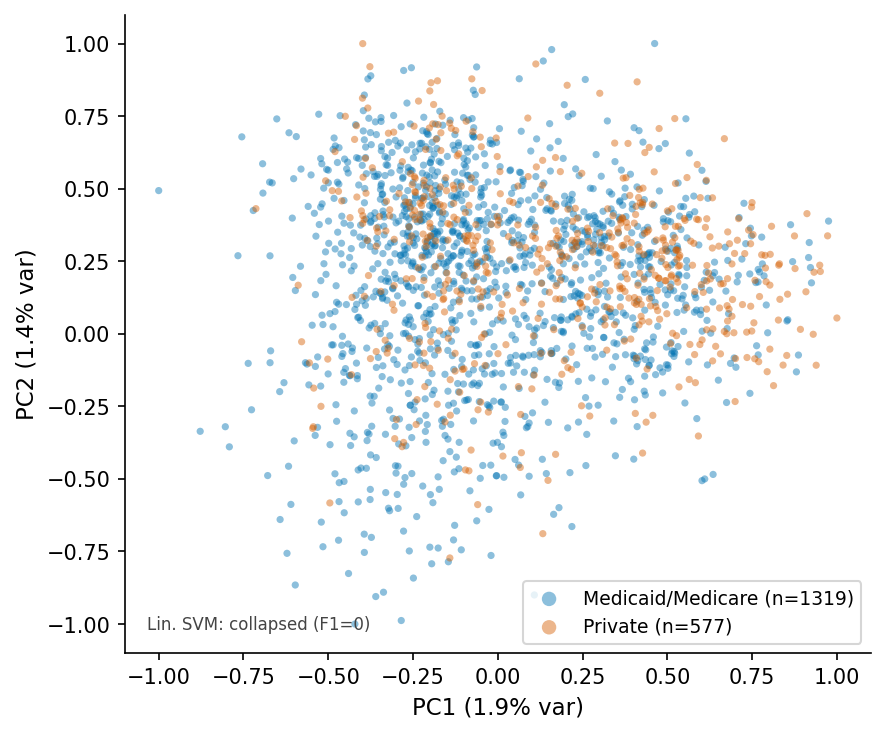}
  \hfill
  \includegraphics[width=0.48\textwidth]{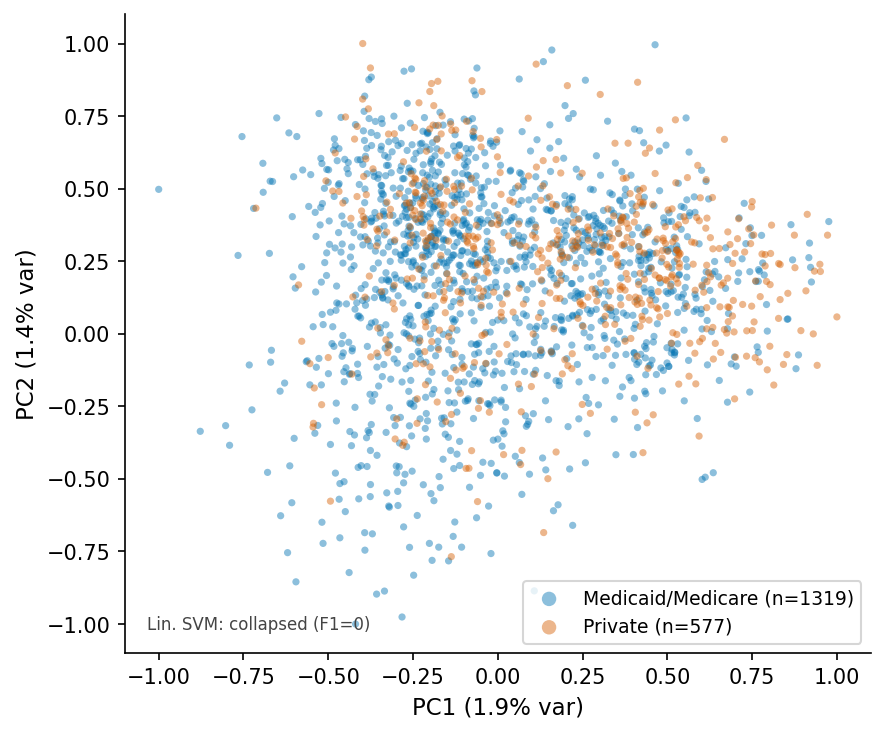}
  \\[4pt]
  \includegraphics[width=0.48\textwidth]{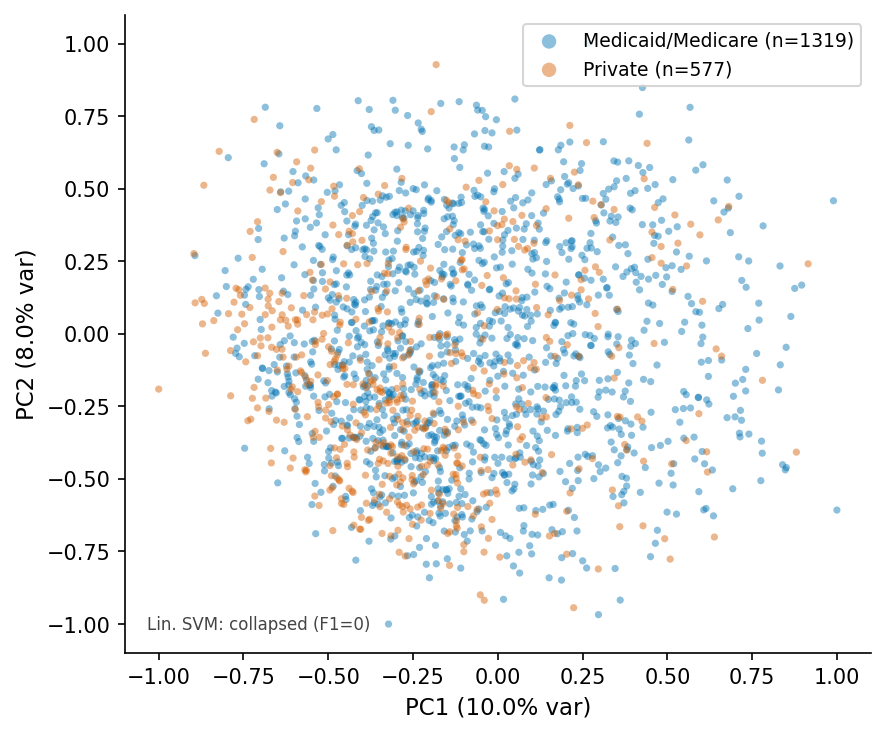}
  \hfill
  \includegraphics[width=0.48\textwidth]{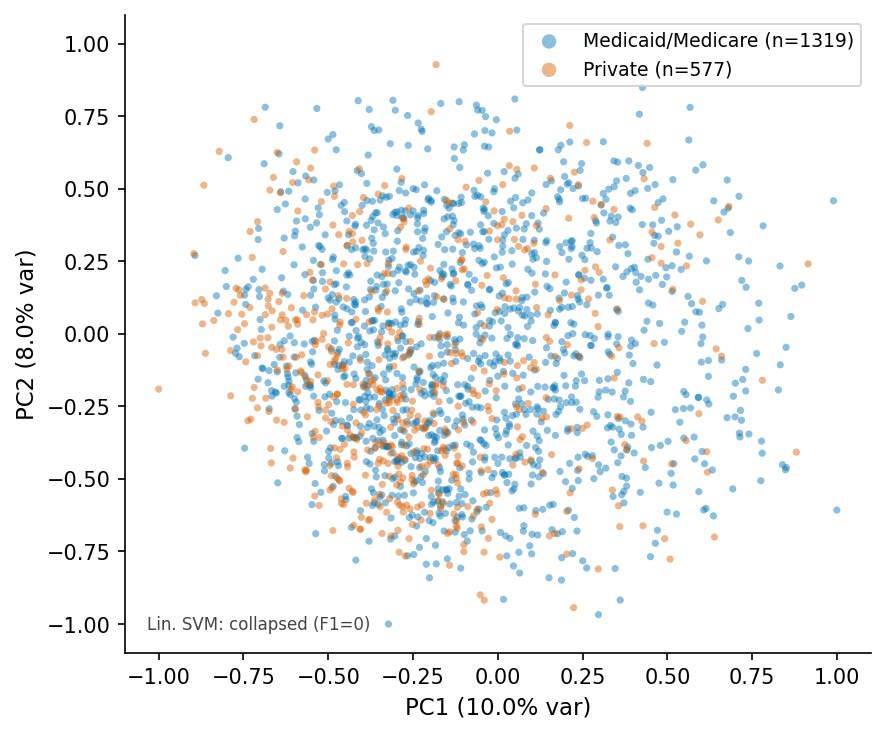}
  \caption{PCA feature space ($q=4$, left; $q=6$, right) for all three models
    (MedSigLIP-448, RAD-DINO, ViT-patch32-cls, top to bottom).
    Each point is a training sample projected onto its first two PCA components after
    StandardScaler$\to$PCA$\to$MinMaxScaler preprocessing (seed~0).
    Blue: Medicaid/Medicare; orange: Private insurance.
    The two classes overlap substantially in every panel, which explains why the linear
    kernel $K_L$---operating in this same $q$-dimensional subspace---collapses to
    majority-class prediction (F1\,=\,0).
    The label in the lower-left corner of each panel confirms classical SVM collapse at
    that configuration.}
  \label{fig:scatter_all}
\end{figure*}

Figure~\ref{fig:scatter_all} shows the PCA-compressed training data at $q=4$ and $q=6$
for all three models. The substantial class overlap visible in every panel provides
a geometric explanation for why the linear kernel collapses.

\subsection{PCA Geometry of MedSigLIP-448 at $q=2$}
\label{sec:supp_pca}

\begin{figure*}[!t]
  \centering
  \includegraphics[width=0.80\textwidth]{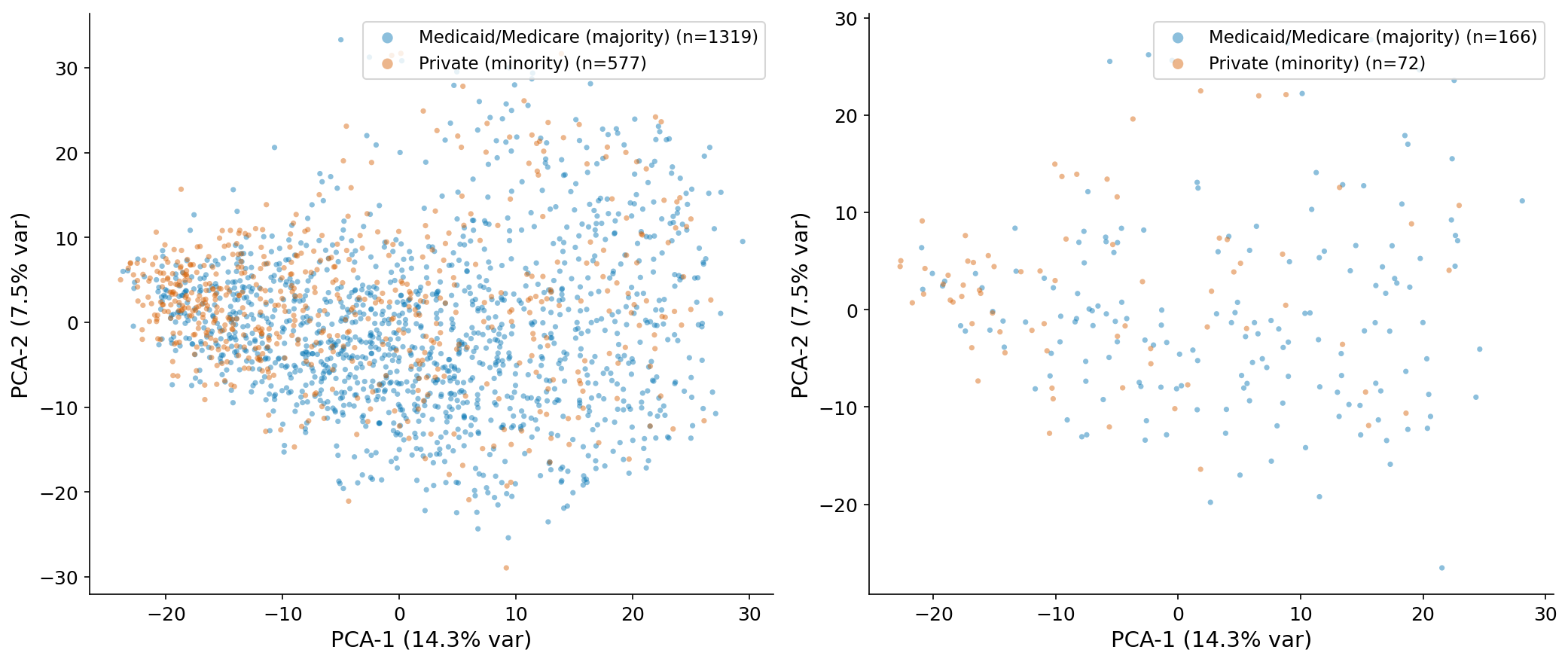}
  \caption{PCA scatter of MedSigLIP-448 embeddings projected to 2 components
    (total explained variance: 21.8\%). Train set: 1319 majority (Medicare)
    vs 577 minority (Private). The low explained variance confirms that the
    2D PCA projection captures only a fraction of the structure exploited
    by the quantum kernel in higher dimensions.}
  \label{fig:pca_scatter}
\end{figure*}

\subsection{ViT-patch32-GAP Pooling Ablation}
\label{sec:supp_gap}

To assess the effect of pooling strategy on quantum kernel performance,
we evaluate a global average pooling (GAP) variant of ViT-patch32
alongside the CLS-token variant reported in the main text.
Both variants produce 768-dimensional embeddings from the same frozen
ViT-patch32 backbone; the only difference is the aggregation of patch tokens:
GAP averages all patch tokens, while CLS uses only the class token.
Multi-seed experiments (10 seeds, DT9, trace normalization, C=1) were
completed for both variants across 11 qubit configurations ($q \in
\{2,3,4,5,6,8,9,10,11,12,16\}$).

Table~\ref{tab:gap_cls_comparison} reports QSVM minority-class F1 (mean $\pm$ std over
10 seeds) for both pooling variants, together with the best classical
SVM baseline (RBF kernel, C=1) evaluated on the same splits.
The two pooling strategies yield nearly identical QSVM performance at
$q \ge 10$ (difference $\le 0.003$), with CLS slightly higher on average
across all $q$.
Both variants show a quantum advantage over the best classical
baseline for $q \ge 4$ under noiseless simulation.
The CLS variant was selected for the main text because it matches the
standard ViT evaluation protocol and shows marginally more consistent F1
across the full qubit sweep.

\begin{table}[h]
\caption{CLS vs.\ GAP pooling: QSVM minority-class F1 (mean $\pm$ std, 10 seeds)
and best classical SVM baseline (RBF, C=1) for ViT-patch32 on DT9.
Both pooling variants produce 768-dimensional embeddings.
$\Delta_\text{GAP}$ = QSVM-GAP $-$ Best-Classical.}
\label{tab:gap_cls_comparison}
\begin{ruledtabular}
\begin{tabular}{c|cc|cc|c}
$q$ & QSVM-CLS & QSVM-GAP & Best-Cl & $\Delta_\text{CLS}$ & $\Delta_\text{GAP}$ \\
\hline
2  & $0.016 \pm 0.028$ & $0.000 \pm 0.000$ & $0.033$ & $-0.018$ & $-0.033$ \\
3  & $0.003 \pm 0.008$ & $0.008 \pm 0.025$ & $0.041$ & $-0.038$ & $-0.018$ \\
4  & $0.048 \pm 0.073$ & $0.102 \pm 0.092$ & $0.044$ & $+0.004$ & $+0.058$ \\
5  & $0.212 \pm 0.101$ & $0.205 \pm 0.110$ & $0.180$ & $+0.032$ & $+0.025$ \\
6  & $0.272 \pm 0.132$ & $0.235 \pm 0.090$ & $0.219$ & $+0.054$ & $+0.016$ \\
8  & $0.370 \pm 0.092$ & $0.278 \pm 0.122$ & $0.251$ & $+0.119$ & $+0.027$ \\
9  & $0.384 \pm 0.081$ & $0.276 \pm 0.141$ & $0.272$ & $+0.112$ & $+0.004$ \\
10 & $0.391 \pm 0.081$ & $0.383 \pm 0.069$ & $0.287$ & $+0.103$ & $+0.096$ \\
11 & $0.403 \pm 0.064$ & $0.406 \pm 0.092$ & $0.303$ & $+0.100$ & $+0.103$ \\
12 & $0.405 \pm 0.063$ & $0.412 \pm 0.082$ & $0.307$ & $+0.098$ & $+0.105$ \\
16 & $0.399 \pm 0.098$ & $0.397 \pm 0.082$ & $0.330$ & $+0.104$ & $+0.067$ \\
\end{tabular}
\end{ruledtabular}
\end{table}

\subsection{ViT-patch16-cls Patch-Size Ablation}
\label{sec:supp_patch16}

To assess the effect of patch size on quantum kernel performance,
we evaluate a ViT with patch size 16 (ViT-patch16-cls,
768-dimensional CLS-token embeddings) alongside the ViT-patch32-cls
variant reported in the main text.
Both variants use the same frozen ViT backbone architecture; the only
difference is the spatial resolution of the patch tokenisation (patch16
produces $4\times$ more tokens per image than patch32 and yields a denser
spatial representation before CLS pooling).
Multi-seed experiments (10 seeds, DT9, trace normalization, C=1) were
completed for ViT-patch16-cls across the same 11 qubit configurations.

Table~\ref{tab:patch16_comparison} shows that ViT-patch16-cls yields
substantially lower QSVM minority-class F1 than ViT-patch32-cls at every qubit
count tested ($\Delta \text{F1} \approx -0.24$ at $q=16$, $-0.28$ at $q=8$).
This performance gap is the primary basis for selecting ViT-patch32-cls
as the main-text ViT baseline.
The result suggests that the denser patch16 representation introduces
additional redundancy or noise in the low-dimensional PCA subspace,
making it harder for the quantum kernel to separate insurance classes.

\begin{table}[h]
\caption{Patch-size ablation: QSVM minority-class F1 (mean $\pm$ std, 10 seeds)
for ViT-patch32-cls and ViT-patch16-cls on DT9.
Both use CLS-token embeddings; best classical SVM baseline (RBF, C=1)
shown for ViT-patch32-cls (the main-text model).}
\label{tab:patch16_comparison}
\begin{ruledtabular}
\begin{tabular}{c|cc|c}
$q$ & QSVM patch32 & QSVM patch16 & $\Delta$ (patch32 $-$ patch16) \\
\hline
2  & $0.016 \pm 0.028$ & $0.000 \pm 0.000$ & $+0.016$ \\
3  & $0.003 \pm 0.008$ & $0.000 \pm 0.000$ & $+0.003$ \\
4  & $0.048 \pm 0.073$ & $0.008 \pm 0.024$ & $+0.040$ \\
5  & $0.212 \pm 0.101$ & $0.005 \pm 0.016$ & $+0.207$ \\
6  & $0.272 \pm 0.132$ & $0.021 \pm 0.031$ & $+0.251$ \\
8  & $0.370 \pm 0.092$ & $0.069 \pm 0.110$ & $+0.301$ \\
9  & $0.384 \pm 0.081$ & $0.079 \pm 0.118$ & $+0.305$ \\
10 & $0.391 \pm 0.081$ & $0.080 \pm 0.148$ & $+0.311$ \\
11 & $0.403 \pm 0.064$ & $0.121 \pm 0.147$ & $+0.282$ \\
12 & $0.405 \pm 0.063$ & $0.104 \pm 0.159$ & $+0.301$ \\
16 & $0.399 \pm 0.098$ & $0.160 \pm 0.118$ & $+0.239$ \\
\end{tabular}
\end{ruledtabular}
\end{table}

% ============================================================
\end{document}